\documentstyle[preprint,aps,floats,epsf,tighten,eqsecnum]{revtex}
\begin{document}
\draft

\preprint{
 \parbox{1.5in}{\leftline{JLAB-THY-97-42}
                \leftline{WM-97-111}}}
\title{Sensitivity of the deuteron form factor to nucleon
resonances$^*$} 
\author{Kelly Ann Herbst}
\address{College of William and Mary, Williamsburg, Virginia 23187}
\author{Franz Gross} 
\address{College of William and Mary, Williamsburg, Virginia 23187 
\protect\\ Thomas Jefferson National Accelerator Facility,
\protect\\12000 Jefferson Avenue, Newport News, Virginia 23606}
\date{\today}
\maketitle

\begin{abstract}

The sensitivity of the deuteron form factor to
contributions from the excited states of the nucleon is explored using a
simple model of the nucleon-nucleon interaction which employs a tower
of charged nucleon resonances.  The  model is manifestly covariant,
analytically solvable, and gauge invariant.  The consequences of this
model are studied in the simplest possible framework.  We assume that
all particles have spin zero and that the tower has only
three charged members, which consist of the proton, the Roper, and a
higher state in the vicinity of the ${D}_{13}$.  Nucleon-nucleon
$S$-wave phase shifts and the deuteron form factor are calculated
using this three member tower, and the results are compared to similar
calculations using the proton ground state only.  We conclude that
the deuteron form factor is insensitive to the presence of excited
states of the proton unless those states are of sufficiently low
mass to produce strong inelasticities in $NN$ scattering channels. 

\end{abstract}

\narrowtext

%
%
\section{Overview, results, and conclusions}
\label{overview}

%
%
\subsection{Introduction}
\label{introduction}

Now that first results from experiments performed at the Thomas 
Jefferson National Accelerator Facility (Jefferson Lab) are becoming
available, the details of the two nucleon current may be explored in
great detail.  Some of the new  experiments will probe the short range
structure of this current, which can be interpreted as arising from the
exchange of quarks and gluons, or alternatively, from meson exchange
and the excitation of virtual excited states of the nucleons.  In
this paper we turn our attention to virtual contributions from these
nucleon excited states.  

The effects of nucleon resonances (or excited states) are obvious
above the production threshold; it is less obvious what effect they
will have on cross sections and observables measured well below the
production threshold.   Denoting the {\it mass\/} of the final
state in electron scattering by $M_x$, and the square of the four
momentum transfer by $Q^2>0$, we try to answer the following question:
``If $M_x$ lies below the threshold for the physical excitation
of nucleon resonances, does their existance nevertheless affect the
high $Q^2$ behavior of form factors and structure functions?''  To
study this question we introduce the ``tower of states'' model.  This
is a simple, dynamical model for the study of nucleon resonances which
can make easy, but admittedly rough, predictions of the effects that
might be seen at places like Jefferson Lab.  The goal is to use this
simple model to get a qualitative feeling of where effects due to
resonances will be large.

Any such model must be compatible with highly successful theories of 
$NN$ interactions \cite{r68}\ --\cite{gvoh92} which have gone before. 
Motivated by previous work \cite{gvoh92} and the relativistic work of
Tjon and collaborators \cite{ft75nc}\ --\cite{ft86} we use manifestly
covariant dynamics.  When working at the level of a few GeV, as at
Jefferson Lab, one must use relativistic equations.  The model is
constructed so that gauge invariance is satisfied exactly.  This
requires the introduction of an interaction current which is
derived from the strong kernel.  Finally, the model is  analytically 
solvable, allowing for simple calculations of various properties of the
$NN$ system.  Many of these features have appeared in earlier models,
but, to our knowledge, this is the first time the contributions from
resonances and their corresponding interaction currents have been 
systematically applied to the study of the deuteron form factor
using a simple model which is analytically solvable, covariant, and
gauge invariant.  

In this first study we use the simplest possible potential 
(rank-one separable), with a minimal number of free parameters (five,
at most).  We fix all free parameters by fitting the $NN$  phase
shifts below lab energies of 350 MeV.  All deuteron properties
(binding energy, wavefunctions, form factors) are then calculated 
from the model.  Our  treatment of the nucleon resonances varies
somewhat from earlier works.  Most models have considered the
addition of only one additional resonance, usually the $\Delta$
(Ref.~\cite{sk80} is an exception, however, but see the discussion in
Ref.~\cite{ft86}).  Such channels have been considered as add-ons to 
existing successful $NN$ models \cite{sp90}, or used to increase the
applicability of such models \cite{ft86}.  Previous calculations of
the deuteron form factor using separable models have usually ignored the
interaction current which should accompany the calculation.  The
philosophy behind our treatment is that the resonances and their
interaction currents should be consistently included on an equal footing
from the beginning.

Since we wish to keep this first exploration of the contribution
of  resonances as simple as possible, we have neglected spin (all
particles in the following discussions are scalar), and used a basic
separable four-point interaction diagram as the kernel with a
``tower'' consisting of only three states.  A Tabakin-style form
factor, similar to the one used by Rupp and Tjon
\cite{rt88}, is employed in order to realistically model the
contributions of meson exchange without actually having to deal with
its complexities.  We compare results from a three member
tower with those obtained from a purely elastic scattering model (a
one member tower, in our language). Therefore, any differences
between the  models which include higher resonances and those which
do not are due to the  presence of the resonances.  Clearly, if
differences between these models can be seen even at this basic
level, then further investigation with more complicated systems is
warranted.

In this section we begin with an overview of the techniques used in
formulating the  model, and then discuss the
results of this work.  Particular attention is paid to the
effects of inelasticity on the tower models.  At the end of this
section we present our conclusions. 
Sec.~\ref{general} includes  a full development of the model,
including the solution of the relativistic scattering equation,
proof of unitarity, a  proof of gauge invariance, and the development
of the deuteron form  factor.   The relations used for inelastic
scattering are also derived.  In order to maintain gauge invariance,
contributions from an interaction current are required, since the
kernel of the this model is dependent on the center-of-mass
four-momentum. The interaction current is fully derived in the Appendix,
and the contributions from the interaction current prove to be very
important to several facets of the $NN$ system.  Finally, Sec.~\ref{tower}
includes some specific details of the models.

%
%
%
\subsection{Background and overview}
\label{background}

The tower of states model is built upon the idea that, in the process 
of repeated scattering with a neutron, a proton may ``sample" any one 
of  its higher order resonances (the ``tower'').  That is, the proton
and neutron interact, causing the proton to become a virtual Roper for a
short period of time  before interacting again with the neutron.  In
successive scatterings, multiple samplings of this tower of higher
resonant states could occur.  We explore the effects that such
sampling will have on various properties of the 
$NN$ system, particularly the properties of the deuteron where all
inelastic contributions are virtual.

\begin{figure}[t]
\begin{center}
\mbox{
   \epsfysize=1.5in
\epsfbox{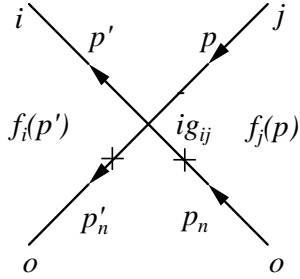}
}
\end{center}
\caption{Basic tower of states interaction.  This diagram forms the 
kernel of the spectator equation describing the scattering and bound
states of the  system.  The cross on the neutron lines denotes that 
the particle is on  the  mass shell.}
\label{kernel}
\end{figure}

As a brief overview of the techniques used in formulating the model, 
we begin with the most general form of the relativistic scattering
equation
\begin{equation}
M = {\cal V} + \int_{k}{\cal V}GM,
\label{basrelscat}
\end{equation}
where ${\cal V}$ is the relativistic kernel, {\it G} is the 
propagator, and $\int_{k}$ is specified by choosing a particular
formalism, such as Bethe-Salpeter \cite{sb51}, Blankenbecler-Sugar
\cite{bs66}, or Spectator \cite{g69}--\cite{g93book}.  The  basic
four-point interaction between a member of the tower and a neutron
shown in  Fig.~\ref{kernel} is chosen as the kernel.  The coupling
strengths are denoted by the coupling matrix $g_{ij}$, where $i$
and $j$ denote members of the tower.  We include a Tabakin-style form
factor 
\begin{equation}
f(x) = \frac{(\alpha^{2} - x) (p_{c}^{2} + x)} 
{(\gamma^{2} - x) (\beta^{2} - x)^{2}},
\label{formfactor}
\end{equation}
with
\begin{equation}
x = x(p,P)={p}^{2} - \frac{(P\cdot {p})^{2}}{P^{2}},
\label{xdef}
\end{equation}
where ${p}$ is the relative momentum between the
neutron and the tower member and $P$ is the center of mass 
four-momentum (for a detailed definition of the form factor, see
Sec.~\ref{specific}). As shown by Rupp and Tjon \cite{rt88}, and
again here for a one-member tower called Model 1,  the Tabakin form
factor allows one to successfully model the $NN$ system without
having to actually perform meson exchange calculations.   Since 
spin is being neglected at this time, all  particles have scalar
propagators.  Finally, we choose to  work in the spectator 
formalism \cite{g69}--\cite{g93book}, which has  proven to be 
very useful in  describing the $NN$ system 
\cite{gvoh92}, \cite{g74}.  For simplicity in this first treatment,
we will treat the $np$ system only, and discard the isospn
formalism, so that the neutron and proton are nonidentical
particles.  The spectator formalism is then simplified if the neutral
particle, i.e.~the neutron, is placed consistently on its mass shell and
excluded from the tower of states.  The towers will contain the
proton and its charged excited states only.  

%

These choices are not made arbitrarily.  Each has been specifically 
chosen to reduce the complexity of the system in some way without
losing the essential physics.  By constructing a very  basic, separable
kernel, the $M$-matrix may be obtained  analytically (see
Sec.~\ref{general} for the full details).  Neglecting spin is, of
course, an automatic reduction of complexity useful for this first
treatment of the problem.  The  choice of the spectator formalism
motivates the initial inclusion of only one tower (since one particle
is on shell, it is natural to restrict it to its ground state, although
a more complete treatment can be developed later) and is chosen because
it is easily extended to more complex cases.  Finally, treating the
proton and neutron as nonidentical particles and neglecting the
distribution of charge inside the neutron allows us to restrict the
electromagnetic interaction to the tower members; the neutron does
not interact with the photon in the naive models considered here.

%
\begin{figure}[t]
\begin{center}
\mbox{
   \epsfysize=1.5in
\epsfbox{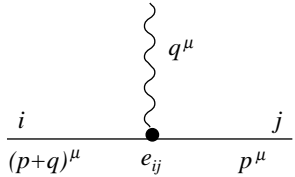}
}
\end{center}
\caption{The basic electromagnetic coupling between tower members}
\label{em}
\end{figure}

Once the scattering equation has been solved, we obtain the
$S$-wave phase shifts, $\delta$, and inelasticity parameters,
$\eta = \cos \rho$, from the usual representation of the inelastic
scattering amplitude 
\begin{equation}
-{M} = \frac{\eta e^{2i\delta _{1}} - 1}{2i\rho_1}\, ,
\label{inelm}
\end{equation}
where $\rho_1$ (not to be confused with the inelastic phase angle
$\rho$) is the relativistic phase space factor given in
Eq.~(\ref{phsspc}) below.  The free parameters of the models are then
fit to the Nijmegen 1993 phase shifts below 350  MeV \cite{n93pc}. 
Excellent fits  are obtained using towers with either one or three 
members.  Fits using a two member tower were attempted, but
proved to be less successful than those obtained from one- and
three-member towers.  The reason for these poor fits is under
investigation; it may be related to the need for a certain symmetry
between the coupling constants.  Due to the larger number of
free parameters, $\chi ^{2}$s for three-member tower models are 
slightly improved over those using only the proton.   These results
will be presented in Sec.~\ref{numerical}.

In order to explore the effect of the resonances in a simple
$NN$ system where they can can only make virtual contributions, we
study the deuteron form factor.  The electromagnetic coupling
between tower members, illustrated in  Fig.~\ref{em}, is chosen
to be proportional to $e_{ij} =\zeta_{ij}e$, where $e$ is the electric
charge.  The full currents with electromagnetic form factors
are constructed using the methods of Gross and Riska \cite{gr87} (see
Eq.~(\ref{currents}) for more detail).  In this simple first
treatment we choose $\zeta_{ij}=\delta_{ij}$ so the photon itself cannot
excite or de-excite members of the tower.   In order to satisfy  gauge
invariance an interaction current must be included, and the inclusion
and study of this current is one of the novel features of this work. 
This current is derived from the kernel ${\cal V}$ by minimal
substitution using the methods of Ito et.~al.~\cite{ibg91}. 
However, because the structure of the kernel is different from the one
assumed by Ito et.~al., a slightly different argument is needed, and
because the derivation is somewhat lengthly the details are presented
in the Appendix. 

%
%
%
\subsection{Numerical results}
\label{numerical}

From the real and imaginary parts of $M$ we calculate the phase
shifts and perform a least-squares fit to  fix the free
parameters in the models.  The resulting parameters for three models
are shown in Table \ref{param}.  Model 1  consists of a tower with the
proton only, Model 3 employs a three-member tower consisting of the
proton, Roper, and ${\rm D}_{13}$, and Model 3I uses the same
three-member tower, except that the mass of the Roper has been
lowered to increase its  inelastic contributions.   The  inelastic
region has no effect on the fitting procedure, as only data below 350
MeV are used in the fit.

%
\begin{table}
\widetext
\squeezetable
\caption{Parameters in the tower of states models.  Numbers in {\bf 
bold  face} were varied during the fitting procedure.  The form factor
parameters and all masses are in GeV; all the coupling constants have
dimensions ${\rm GeV}^{4}$.}
\label{param}
\begin{tabular}{ccccccc}
 & \multicolumn{2}{c}{Model 1} & \multicolumn{2}{c}{Model 3} & 
\multicolumn{2}{c}{Model 3I} \\ Parameter & $^{1}S_{0}$ & $^{3}S_{1}$ 
&
$^{1}S_{0}$ & $^{3}S_{1}$ & $^{1}S_{0}$ & $^{3}S_{1}$ \\
\tableline
$\alpha$ & {\bf 0.198125} & {\bf 0.155094} & {\bf 0.197611} &  {\bf
0.157080} & {\bf 0.197750} & {\bf 0.150054} \\
$\beta$ & {\bf 1.09772} & {\bf 1.27872} & {\bf 1.09795} &  {\bf
1.29825} & {\bf 1.10168} & {\bf 1.28994} \\
$\gamma$ & {\bf 0.171623} & {\bf 0.136404} & {\bf 0.172138} &  {\bf
0.139047} & {\bf 0.172141} & {\bf 0.129887} \\
$p_{c}$ & 0.3492 & 0.412689 & 0.3492 & 0.412689 & 0.3492 & 0.412689 \\
$g_{11}$ & -1516.96 & -2553.31 & {\bf -1500.89} & -2615.86 & {\bf 
-1504.81} & -2554.04 \\
$g_{12}$ & & & 155.0 & {\bf 406.731} & 190.0 & {\bf 404.563} \\
$g_{13}$ & & & 155.0 & -105.0 & 190.0 & -105.0 \\
$g_{22}$ & & & 10.0 & -105.0 & 15.0 & -105.0 \\
$g_{23}$ & & & {\bf 272.175} & {\bf 597.206} & {\bf 253.928} & {\bf 
591.489} \\
$g_{33}$ & & & 10.0 & -105.0 & 15.0 & -105.0 \\
$\chi^{2}$ & 6.97 & 7.60 & 5.06 & 6.83 & 3.66 & 6.04 \\
$m_{1}$ & \multicolumn{2}{c}{0.93825} & \multicolumn{2}{c}{0.93825} & 
\multicolumn{2}{c}{0.93825} \\
$m_{2}$ & \multicolumn{2}{c}{ } & \multicolumn{2}{c}{1.44} & 
\multicolumn{2}{c}{1.17} \\
$m_{3}$ & \multicolumn{2}{c}{ } & \multicolumn{2}{c}{1.52} & 
\multicolumn{2}{c}{1.52} \\
\end{tabular}
\narrowtext
\end{table}

%
%
\begin{table}[b]
\caption{Binding energy of the deuteron as calculated in each of the 
three models.  The percent error shown is the error between the
calculated value and the expected value of 2.22 MeV.  The energies are 
given in MeV.}
\label{massdeut}
\begin{tabular}{lccc}
 & Model 1 & Model 3 & Model 3I \\
\tableline
Binding Energy & 2.15 & 2.02 & 2.47 \\
Percent Error & 3.2 & 9.0 & 11.3 \\
\end{tabular}
\end{table}
%

%
%
\begin{table}
\widetext
\squeezetable
\caption{Values for the phase shifts, in degrees, below 350 MeV. 
$T_{{\rm LAB}}$ is in MeV.}
\label{fitphase}
\begin{tabular}{ddddddddd}
 & \multicolumn{2}{c}{Nijmegen 1993\tablenote{Reference \cite{n93pc}}} 
& \multicolumn{2}{c}{Model 1} & \multicolumn{2}{c}{Model 3} &
\multicolumn{2}{c}{Model 3I} \\
$T_{{\rm LAB}}$ & $ ^{1}S_{0}$ & $^{3}S_{1}$ & $ ^{1}S_{0}$ & 
$^{3}S_{1}$ & $ ^{1}S_{0}$ & $^{3}S_{1}$ & $ ^{1}S_{0}$ & $^{3}S_{1}$ 
\\  \tableline
1 & 62.029 & 147.75 & 61.559 & 147.97 & 60.889 & 147.58 & 61.2 &
148.36 \\ 
5 & 63.627 & 118.18 & 64.203 & 118.27 & 63.99 & 117.83 &
64.148 & 118.54 \\ 
10 & 59.956 & 102.61 & 60.55 & 102.32 & 60.498 &
101.96 & 60.614 & 102.35 \\ 
25 & 50.903 & 80.629 & 51.199 & 80.009 &
51.32 & 79.786 & 51.403 & 79.705 \\ 
50 & 40.545 & 62.767 & 40.675 &
62.641 & 40.884 & 62.468 & 40.961 & 62.222 \\ 
75 & 32.933 & 51.585 &
32.912 & 51.852 & 33.149 & 51.685 & 33.232 & 51.435 \\ 
100 & 26.783 &
43.227 & 26.564 & 43.72 & 26.812 & 43.556 & 26.903 & 43.332 \\ 
125 &
21.549 & 36.46 & 21.174 & 37.146 & 21.423 & 36.985 & 21.525 & 36.793 \\
150 & 16.939 & 30.72 & 16.371 & 31.480 & 16.618 & 31.327 & 16.73 &
31.165 \\ 
175 & 12.776 & 25.701 & 12.075 & 26.536 & 12.317 & 26.394 &
12.439 & 26.258 \\ 
200 & 8.943 & 21.216 & 8.08 & 22.026 & 8.3152 &
21.898 & 8.4471 & 21.786 \\ 
225 & 5.357 & 17.14 & 4.4249 & 17.959 &
4.648 & 17.847 & 4.7894 & 17.755 \\ 
250 & 1.959 & 13.386 & 1.031 &
14.189 & 1.2076 & 14.095 & 1.3552 & 14.022 \\ 
275 & -1.3 & 9.89 &
-2.3423 & 10.696 & -2.0353 & 10.621 & -1.8673 & 10.565 \\ 
300 & 
-4.457 & 6.602 & -5.3305 & 7.4381 & -5.0731 & 7.3836 & -4.897 & 7.3436
\\ 
325 & -7.545 & 3.484 & -8.2173 & 4.958 & -7.9706 & 4.3246 &
-7.7858 & 4.3003
\\ 
350 & -10.59 & 0.502 & -10.96 & 1.4582 & -10.72 & 1.4466 & 
-10.525 & 1.438 \\
\end{tabular}
\narrowtext
\end{table}

In order to best understand the nature of the restrictions placed on 
these  parameters, look at the expression for $\tan \delta$.  For a
one-member tower (Model 1), this expression reduces to
\begin{equation}
\tan \delta _{1} = -\frac{g_{11}{\rm Im}{\cal G}_{1}}{4\pi ^{2} + 
g_{11}{\rm Re}{\cal G}_{1}}\, ,
\label{omtdelta}
\end{equation}
where ${\cal G}_1$ is the integral given in Eq.~(\ref{intgdef}). 
In order for the phase shift to cross zero (which it does at around
300 MeV lab energy), there must be a zero in  the numerator, and ${\rm
Im}{\cal G}_{1}$ must be set equal to zero at the particular value of
the center of mass energy at which  the  phase shift crosses zero.  If
we fix
$p_{c}$ by this method, the zero in the numerator will be a double
zero, and will result in the desired structure for the phase shift only
if the denominator also has a single zero located at precisely the same 
center of mass energy.  This zero in the denominator of tan $\delta
_{1}$ is used to determine the parameter $g_{11}$.  Note that the
value of $p_{c}$ will be a constant during the fitting procedure (it
does not  depend on any  of the free parameters) and $g_{11}$ will
be determined actively during the fitting process.  Of the five
parameters which define Model 1, only $\alpha$,
$\beta$, and $\gamma$ are are varied freely during the fit, and these
are shown in {\bf bold face} in Table \ref{param}.  This procedure was
used for both the singlet and triplet cases. 

The careful reader will have noticed that the denominator of  $\tan
\delta _{1}$ in  Eq.~(\ref{omtdelta}) must be the
same as the denominator of the $M$-matrix.  Thus, this expression 
must also have
a second zero in the triplet case, located at the mass of the deuteron. 
No attempt is made to fix the  location of this zero, we instead
determine it  after the fitting is complete, obtaining the mass of the
deuteron directly from the model.
Table \ref{massdeut} shows the values of the binding energy of the
deuteron for the various models, and the percent difference from the
experimental binding energy of the deuteron.  While the these errors are
large on the scale of the deuteron binding energy, 2.2 MeV, they result
from a delicate cancellation between two large numbers (of the order of
2 GeV), and hence the agreement is really quite good.   In light of
this, all three models  give quite satisfactory values for the
binding energy.

%
\begin{figure}[b]
\begin{minipage}{4.0in}
\begin{center}
\mbox{
   \epsfxsize=3.5in
\epsfbox{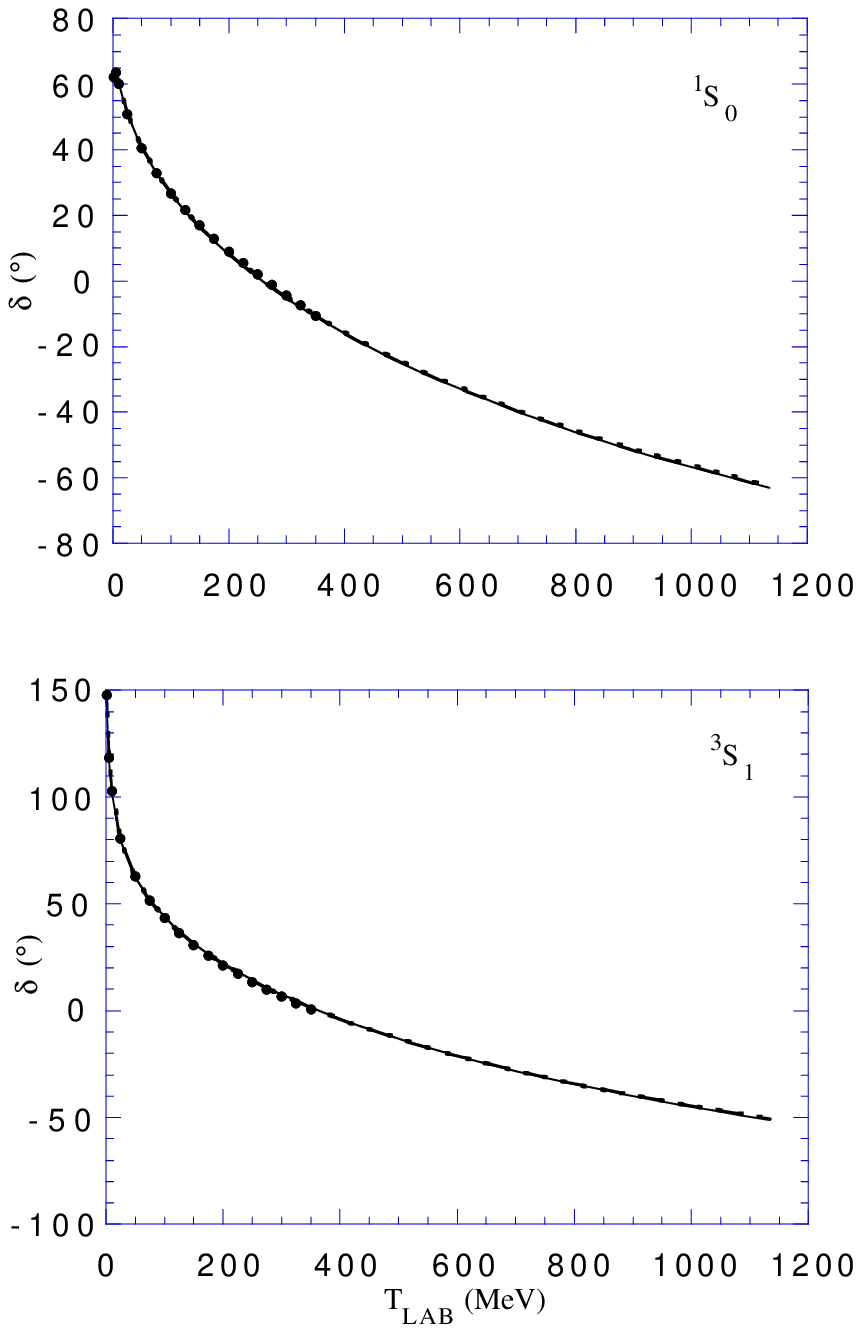}
}
\end{center}
\caption{The singlet and triplet phase shifts to 1.2 GeV in lab energy
for Model 1 (solid line), Model 3 (dashed line), and Model 3I (dotted
line).  The Nijmegen 1993 phase shift data are shown up to 350 MeV 
(Ref.~21) (black circles).  Note that the phase shifts in both cases
are nearly indistinguishable.}
\label{fullphase}
\end{minipage}
\end{figure}

In the case of Models 3 and 3I there are ten parameters: six 
coupling  constants and four form factor parameters.  Unlike the case
of Model 1, the singlet and triplet couplings have different
restrictions placed upon them.   For the singlet case, $p_{c}$
will be fixed using the location of the zero in the phase shift, as 
detailed   above, reducing the number of parameters to nine.  It was
determined that $g_{11}$ needs to be a fitted parameter in this case,
in order to reduce $\chi ^{2}$ to its optimum value.  There are,
however, restrictions on several of the other coupling constants,
specifically, $g_{12} = g_{13}$ and $g_{22} = g_{33}$.  These four
couplings are  then fixed to constant values, reducing the number of
parameters to five. 

The choice of values  for certain coupling constants may seem
arbitrary.  It was discovered during initial fitting trials that the
fits were most sensitive to the values of the form factor parameters,
and rather less sensitive to  most of the tower couplings. 
Therefore, many of the tower couplings were fixed at historically
useful values, preserving certain symmetries  which seemed to
consistently appear in successful fits. These values, for both the
form factor parameters and the tower couplings, are given in Table
\ref{param}.    While the fitting  procedure is most sensitive to the
values of the form factor parameters,  it also shows some sensitivity
to the values of $g_{11}$ and $g_{23}$.  These parameters were
therefore allowed to vary, while the other couplings were fixed such
that the proportions of the couplings were maintained.  For the
triplet case, the full set of restrictions  on
$p_{c}$ and $g_{11}$ is once again applied; only $g_{12}$ and
$g_{23}$ are allowed  to vary.  Once again, the restriction placed on
the remaining coupling constants, $g_{13} = g_{22} = g_{33}$, is
designed to maintain the proportions of the couplings.  Thus, there
are still only five free parameters in the triplet case, but they are
different from those in the singlet case.

We now present numerical results for all three models discussed
above,  as well  as our analysis of the significance of these results. 
First are the  phase shifts, shown in Table \ref{fitphase} for the 
fitting region (below 350 MeV) for the three models together
with the Nijmegen 1993 values.  All three models give nearly
indistinguishable fits to the Nijmegen phase shifts, with  excellent
$\chi ^{2}$s.  Clearly, there is no significant difference  between
the models as regards the phase shifts below 350 MeV.  This is as
expected;  substantial changes in the phase shifts are not 
anticipated until one  goes above the inelastic threshold.

%
\begin{figure}[b]
\begin{minipage}{4.0in}
\begin{center}
\mbox{
   \epsfxsize=3.5in
\epsfbox{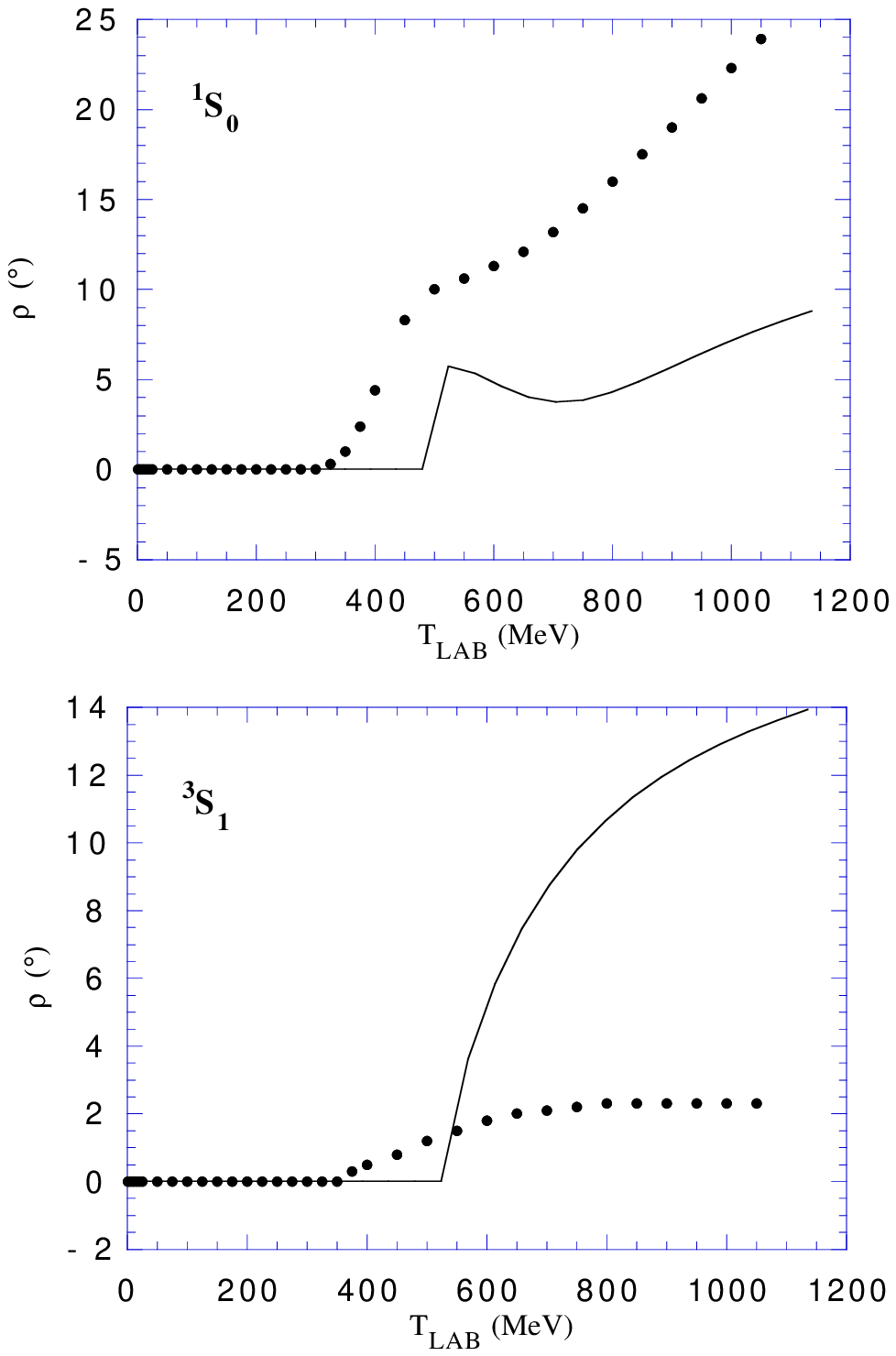}
}
\end{center}
\caption{Inelasticities for Model 3I.  The shape of the $^{1}S_{0}$
inelasticity, while more pronounced than others, is not unusual.  The
black dots are the inleasticities of Arndt and  collaborators 
(Ref.~24).}
\label{rhots}
\end{minipage}
\end{figure}

%
\begin{figure}[t]
\begin{minipage}{4.0in}
\begin{center}
\mbox{
   \epsfxsize=3.5in
\epsfbox{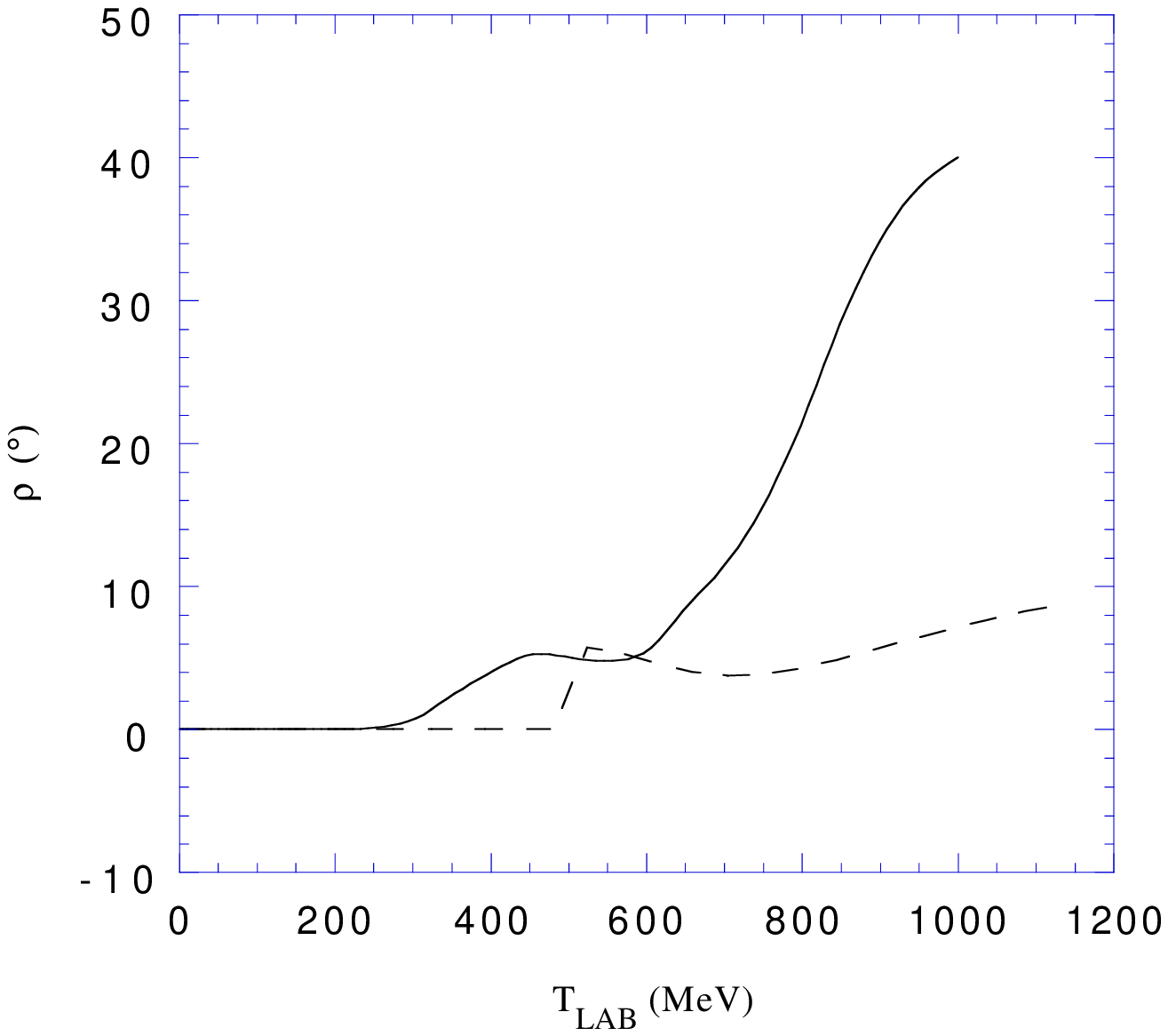}
}
\end{center}
\caption{The $^{1}S_{0}$ inelasticities of van Faassen and Tjon
(Ref.~12) (solid curve) and  the Model 3I (dashed curve).
Note that the shape of the Model 3I inelasticity is similar to that 
of  van Faassen and Tjon, though more pronounced.}
\label{rhoaandt}
\end{minipage}
\end{figure}

After the parameters are determined, the singlet and triplet phase
shifts are calculated out to 1.2 GeV in lab energy and are shown in
Fig.~\ref{fullphase} (the  Nijmegen 1993 phases are also shown in the
fitting region).  All three models fit the phase shifts extremely
well.  More importantly, note that the phase shifts for all three
models are nearly identical in the entire range,  despite the  fact
that the onset of inelasticity in Model 3I is at 492 MeV.  The
inelasticities for Model 3I are shown in Fig.~\ref{rhots}, along with
the inelasticities of Arndt  and collaborators
\cite{arbcvws83}.  We have chosen to plot $\rho$, which is
defined by $\eta = \cos \rho$.  Some readers may be puzzled by the 
shape of the singlet inelasticity.  In Fig.~\ref{rhoaandt} we show
the  inelasticities for the $^{1}S_{0}$ {\em np} partial wave of van
Faassen and Tjon \cite{ft86} together which those of our  Model 3I. 
Ours is more pronounced than that of van Faassen and Tjon, but
the structure is similar.  Some deviation in both the magnitude and
shape  of the inelasticity is expected since the source of
inelasticity in our model arises from the production of a zero-width
nucleon resonance like the Roper, instead of the more realistic 
coupling to a continum $\pi NN$ channel associated with the
onset of production of the $\Delta$ resonance.

%
\begin{figure}
\begin{minipage}{4.0in}
\vspace*{-0.3in}
\begin{center}
\mbox{
   \epsfxsize=3.5in
\epsfbox{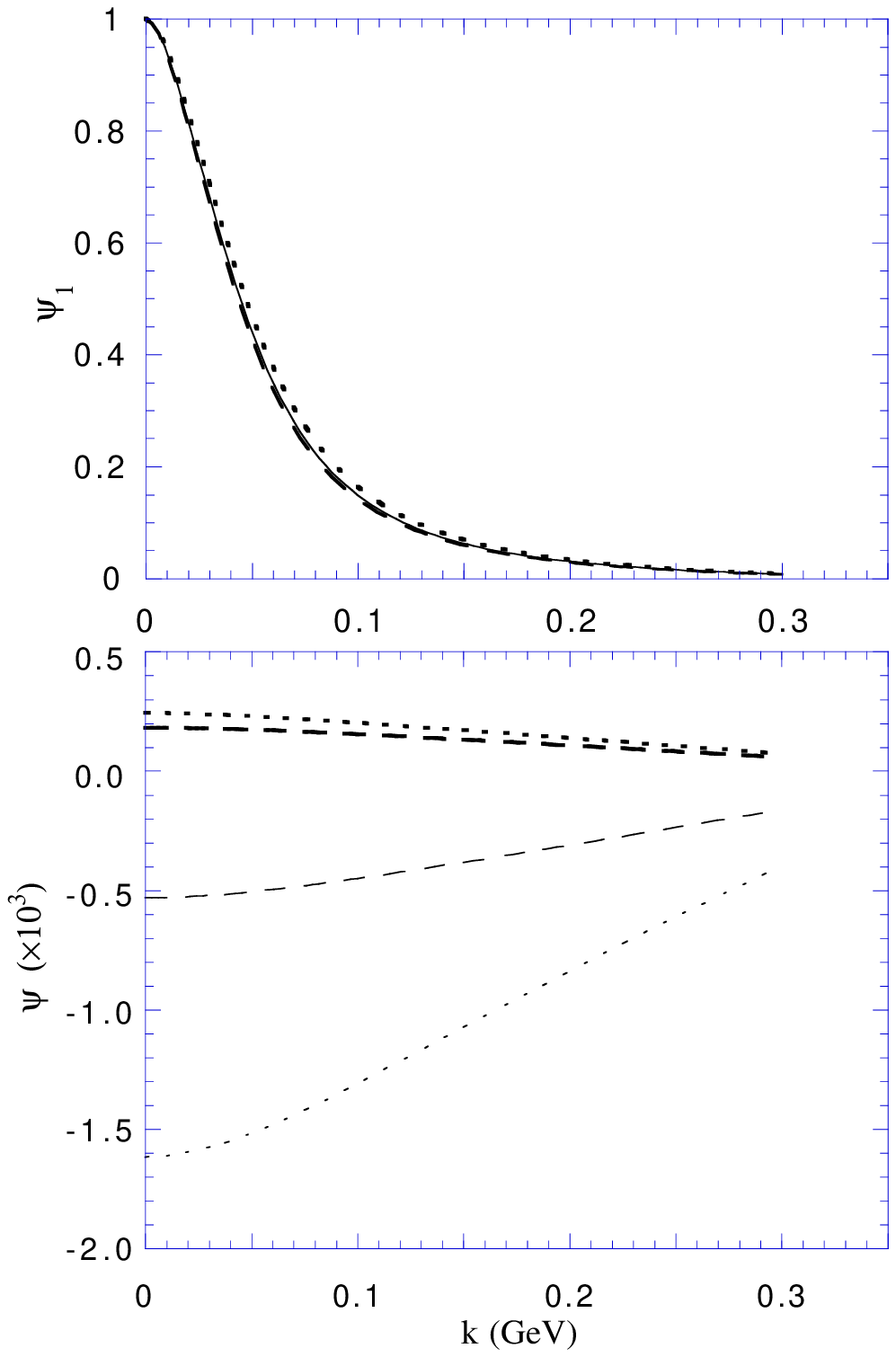}
}
\end{center}
\caption{Deuteron wave functions for Model 1 (solid curves), 
Model 3 (dashed curves),
and Model 3I (dotted curves).  In the lower graph, the {\bf bold face} 
curves are the $\psi_{2}$  components, while the normal curves are 
the $\psi_{3}$ components.  These wave functions are dimensionless 
due to the manner in which they are normalized.}
\label{wavefunctions}
\end{minipage}
\end{figure}

%
\begin{figure}[t]
\begin{minipage}{4.0in}
\begin{center}
\mbox{
   \epsfxsize=3.8in
\epsfbox{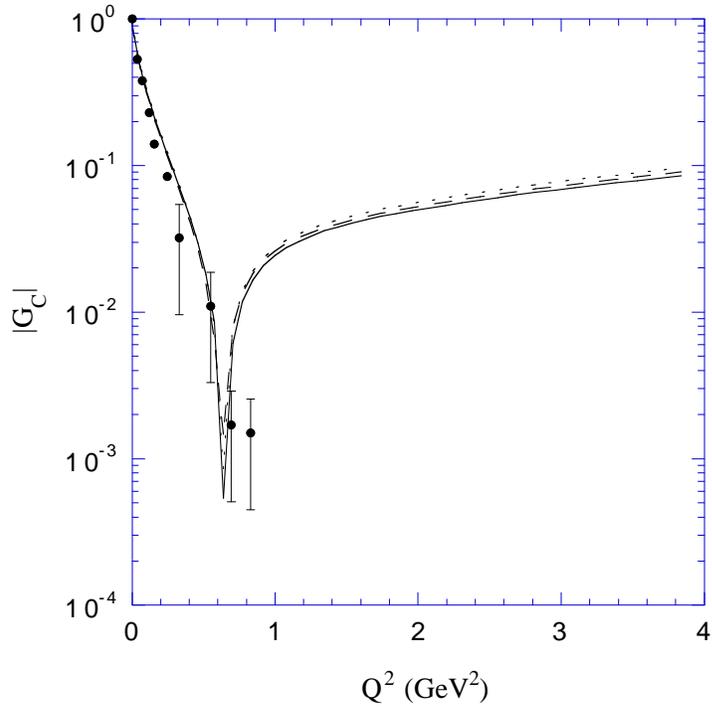}
}
\end{center}
\caption{The deuteron form factor for the three 
models.  The curves are as in Fig.~6. The black
circles are data from Ref.~25.}
\label{dff}
\end{minipage}
\end{figure}
%
\begin{figure}[b]
\begin{minipage}{4.0in}
\begin{center}
\mbox{
   \epsfxsize=3.8in
\epsfbox{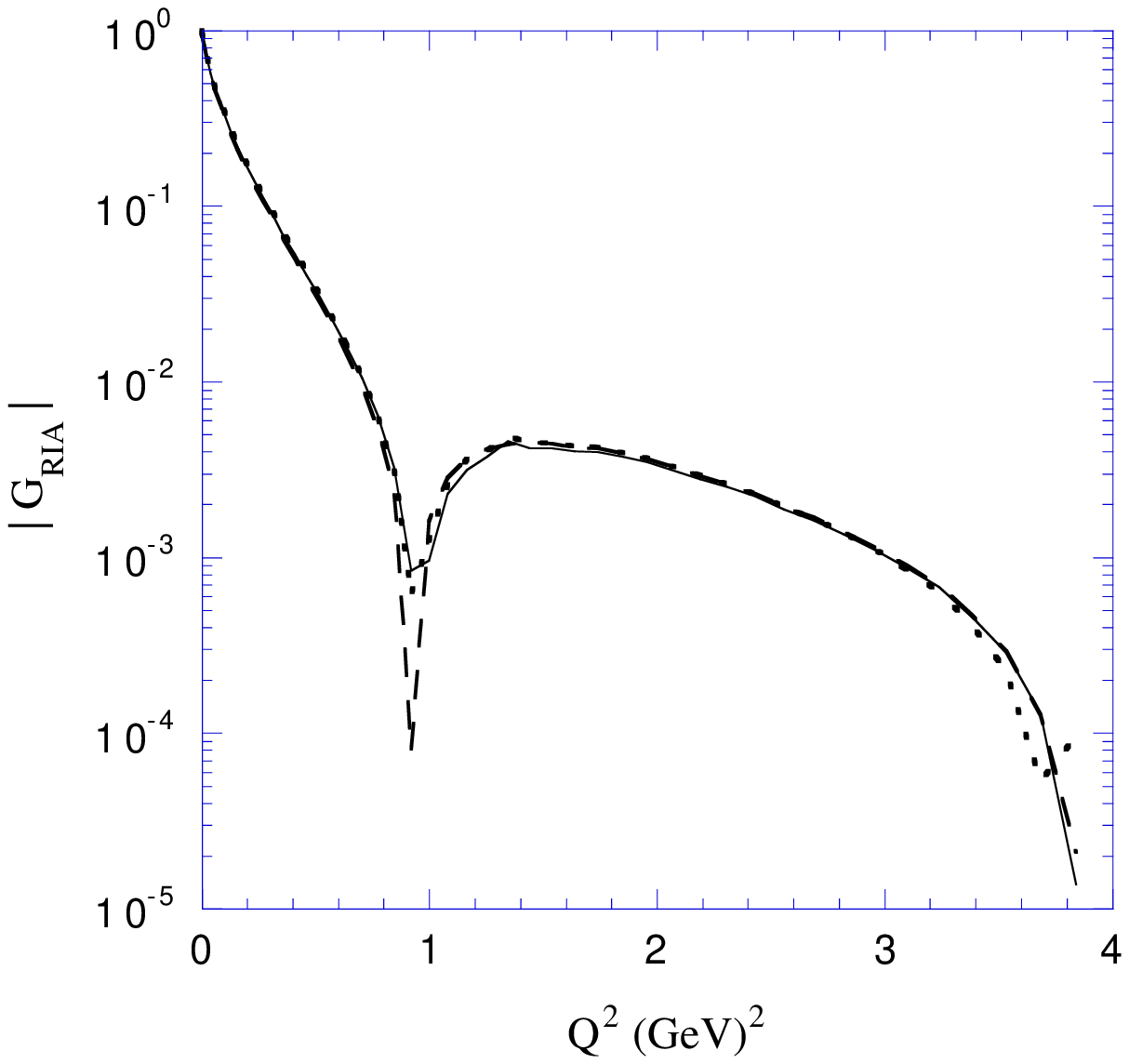}
}
\end{center}
\caption{The relativistic impulse approximation contribution to the
deuteron form factor.  The curves are as in Fig. 7.  Note the presence 
of an additional minimum for Model 3I.}
\label{riacomp}
\end{minipage}
\end{figure}

We have also plotted the deuteron wave function for each of the three 
models in Fig.~\ref{wavefunctions}.  All the wave functions are
normalized in the  following manner:
\begin{equation}
\psi_{i}(k) = \frac{\phi_{i}^{B}(k)}{\phi_{1}^{B}(0)}.
\label{boundpsi}
\end{equation}
Clearly, the contributions from $\psi_{2}$ and $\psi_{3}$ are very 
small indeed. 
 
Finally, we examine the deuteron form factor, shown in Fig.~\ref{dff}.
Note that, in spite of the unrealistic simplicity of the models, all
three of them describe the data \cite{vopc} resonably well.  We
attribute this to the fact that each has been fitted to the $^3S_1$
phase shifts.   There is a small gap between the three curves most
prominently seen at high $Q^2$, where the  average percent difference
between Model 1 and Model 3I is 12\%.  This  effect can only be due
to the presence of the additional resonances in the  three-member
models.  A small shift between Models 3 and 3I can also be  seen.

The separation of the curves in Fig.~\ref{dff}, as well as the
reasonable comparison to the data, appears to be due  entirely to the
presence of the interaction current.  Fig.~\ref{riacomp} shows the
relativistic impulse  approximation contributions for the three
models, which we denote by $F_I$.  We see that they are nearly
identical, except for the  additional minimum just visible near 4 
${\rm GeV}^{2}$ in Model 3I, and would not fit the data alone.
We emphasize that only the diagonal terms of the interaction current
have been included in this calculation of the deuteron form factor
(recall that $\zeta_{ij}=\delta_{ij}$).  Inclusion of the off-diagonal
terms would likely increase the interaction current contribution, and
change the  final result for the deuteron form factor.

Clearly there is an effect due to the presence
of the resonances in the three-member tower models, most easily seen in the 
calculation of the deuteron form factor, but even there it is rather 
small.  However, the question must be  asked, are the effects truly
small, or are they small in this case because the inelasticities
themselves are small?  It is with this idea in mind that, in the next
section,  the effects of large inelasticity on the tower model shall be
explored.

%
%
\subsection{An exploration of large inelasticity}
\label{exploration}

As seen in the previous section, Model 3I has a relatively small
inelasticity (but bigger that the actual data, as shown in
Fig.~\ref{rhots}), yet there seems to be an effect visible in the
deuteron form factor.   We now explore  this effect, specifically
to discover if greater inelasticity will produce larger effects
in such calculated quantities.  A criticism frequently leveled at
current models is that the inelasticities cannot be ignored as one 
proceeds to larger $Q^{2}$.  How large do the inelasticities need to be
before one can no longer ignore their effects?  It is known that in
other partial waves the true $NN$ system has larger inelasticities than
those found for this model, thus it seems reasonable to explore
regions of large inelasticity.

In order to accomplish this, we choose to work backwards, in a sense.  
Rather than try to fit the known inelasticities of the $NN$ system 
with a tower model, we chose a simpler course of action, in keeping
with a first exploration of these concepts.  First, a set of data was
created, based on Model 3I, with much larger inelasticity. 
Then, a one-member tower model, similar to Model 1,  was fit to the
low energy data of this new data set.  The full set of calculations
were then completed for both the new data and the new model.  This
gives us insight into the question of large inelasticity in the
simplest and most direct way possible.
  
The large inelasticity data set was created by adjusting the Roper 
couplings of Model 3I, that is, $g_{12}$, $g_{22}$, and $g_{23}$, such
that the  inelasticity $\rho$ increased to 40-50 degrees, while
maintaining the shape of the phase shifts as much as possible.  The
restrictions placed on the couplings in Sec.~\ref{numerical} are
also kept intact.  The parameters for this set of  data,
called the LI data set, can be found in Table
\ref{arparam}.  This table also includes the parameters of the 
one tower model, referred to as Model 1LI, which was fit to this
data set. The phase shifts for both the LI data set and Model 1LI
are given in Table \ref{arfitphase}.  Note that the changes in the
phase shift are very slight for the singlet case, but somewhat more
pronounced in the triplet (see Table \ref{fitphase} for the Nijmegen
1993 phase shift data).  Even though the $\chi^{2}$
for the triplet fit of Model 1LI seems quite large, the fit is quite
good, as we will see below.  The binding energy of the deuteron for both
the data set and the model are given in Table \ref{armassdeut}.

%
%
\begin{table}
\squeezetable
\caption{Parameters in the LI data set and Model 1LI.  Numbers in
{\bf bold face} were varied during the fitting procedure.  The form
factor parameters and all masses are in GeV; all the coupling
constants have dimensions ${\rm GeV}^{4}$.}
\label{arparam}
\begin{tabular}{ccccc}
 & \multicolumn{2}{c}{LI Data Set} & \multicolumn{2}{c}{Model 1LI} \\
Parameter & $^{1}S_{0}$ & $^{3}S_{1}$ & $^{1}S_{0}$ & $^{3}S_{1}$ \\
\tableline
$\alpha$ & 0.197750 & 0.150054 & {\bf 0.197579} & {\bf 0.075051} \\
$\beta$ & 1.10168 & 1.28994 & {\bf 1.10175} & {\bf 1.31118} \\
$\gamma$ & 0.172141 & 0.129887 & {\bf 0.172310} & {\bf 0.058073} \\
$p_{c}$ & 0.3492 & 0.412689 & 0.3492 & 0.412689 \\
$g_{11}$ & -1511.44 & -1906.51 & -1534.46 & -2758.70 \\
$g_{12}$ & 200.0 & 1000.0 & & \\
$g_{13}$ & 200.0 & -600.0 & & \\
$g_{22}$ & -60.0 & -600.0 & & \\
$g_{23}$ & 1700.0 & 800.0 & & \\
$g_{33}$ & -60.0 & -600.0 & & \\
$\chi ^{2}$ & & & 0.029 & 31.60 \\
$m_{1}$ & \multicolumn{2}{c}{0.93825} & \multicolumn{2}{c}{0.93825} \\
$m_{2}$ & \multicolumn{2}{c}{1.17} & & \\
$m_{3}$ & \multicolumn{2}{c}{1.52} & & \\
\end{tabular}
\end{table}
%

%
%
\begin{table}
\caption{Values for the phase shifts below 350 MeV for the LI data set 
and Model 1LI.}
\label{arfitphase}
\begin{tabular}{ddddd}
$T_{{\rm LAB}}$ & \multicolumn{2}{c}{LI Data Set} & \multicolumn{2}{c}
{Model 1LI} \\
 & $^{1}S_{0}$ & $^{3}S_{1}$ & $^{1}S_{0}$ & $^{3}S_{1}$ \\
\tableline
1.0 & 55.63 & 140.6 & 55.58 & 141.2 \\
5.0 & 61.25 & 109.1 & 61.26 & 108.3 \\
10.0 & 58.52 & 93.78 & 58.56 & 92.78 \\
25.0 & 50.06 & 73.11 & 50.13 & 73.44 \\
50.0 & 40.01 & 57.21 & 40.08 & 58.82 \\
75.0 & 32.46 & 47.31 & 32.53 & 49.36 \\
100.0 & 26.24 & 39.84 & 26.30 & 41.94 \\
125.0 & 20.94 & 33.78 & 20.99 & 35.86 \\
150.0 & 16.2 & 28.57 & 16.24 & 30.51 \\
175.0 & 11.96 & 24.02 & 11.99 & 25.79 \\
200.0 & 8.003 & 19.89 & 8.025 & 21.45 \\
225.0 & 4.384 & 16.17 & 4.397 & 17.52 \\
250.0 & 1.022 & 12.73 & 1.025 & 13.86 \\
275.0 & -2.322 & 9.559 & -2.329 & 10.46 \\
300.0 & -5.284 & 6.62 & -5.301 & 7.277 \\
325.0 & -8.146 & 3.86 & -8.174 & 4.266 \\
350.0 & -10.87 & 1.284 & -10.91 & 1.428 \\
\end{tabular}
\end{table}

\begin{figure}[t]
\begin{minipage}{4.0in}
\begin{center}
\mbox{
   \epsfxsize=3.5in
\epsfbox{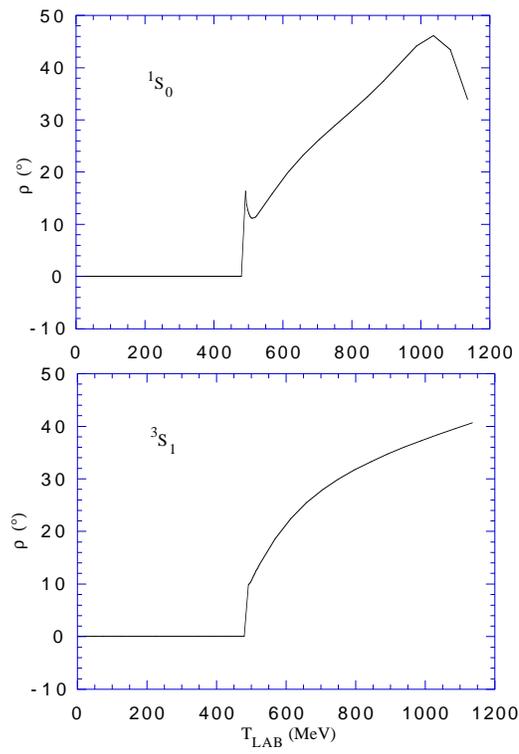}
}
\end{center}
\caption{The singlet and triplet inelasticities for the LI data set.}
\label{arinel}
\end{minipage}
\end{figure}

\begin{figure}
\begin{minipage}{4.0in}
\begin{center}
\mbox{
   \epsfxsize=3.5in
\epsfbox{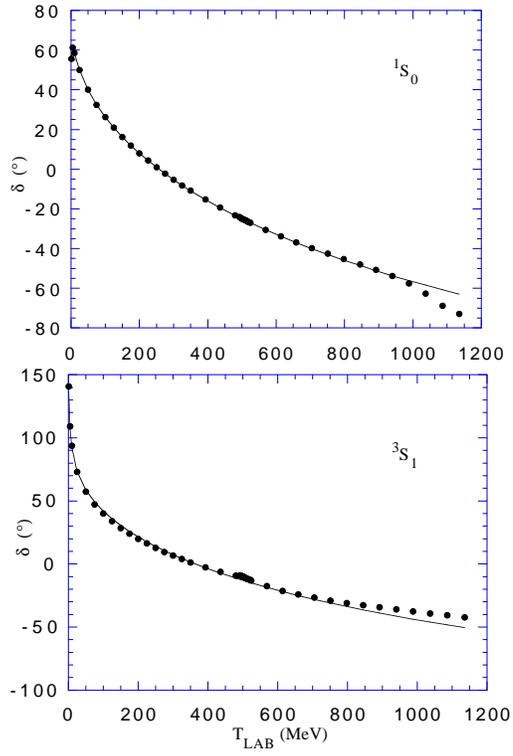}
}
\end{center}
\caption{The phase shifts to 1.2 GeV for the Model 1LI (solid line) and
the  LI data set (black circles).  Note 
that the kink in the triplet phase shift for the LI data set occurs
exactly at the inelastic threshold.}
\label{arfullphase}
\end{minipage}
\end{figure}

\begin{figure}[t]
\begin{minipage}{4.0in}
\begin{center}
\mbox{
   \epsfxsize=3.5in
\epsfbox{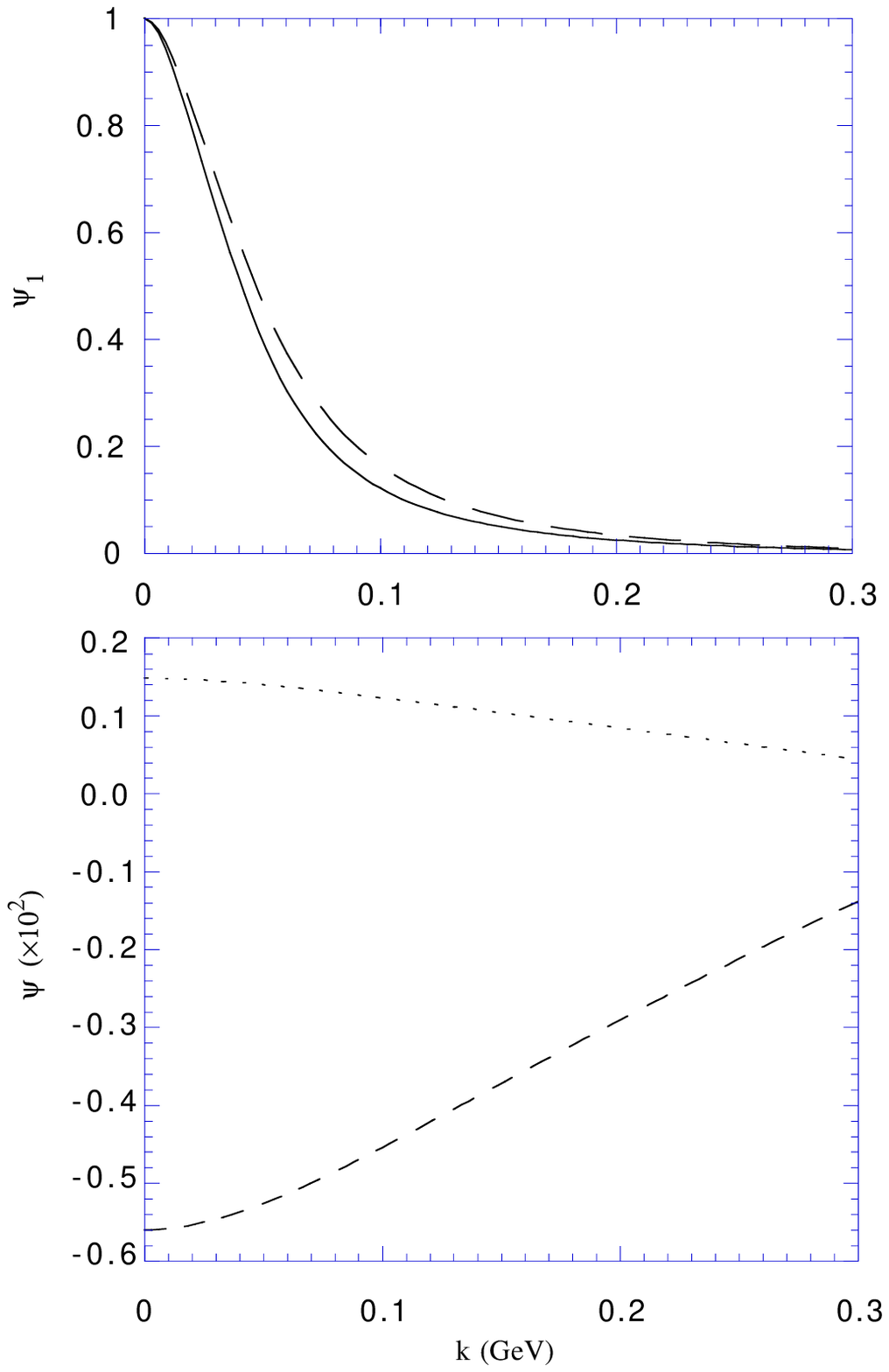}
}
\end{center}
\caption{Deuteron wave functions for Model 1LI (solid curves) and 
LI data (broken curves).  In the lower graph, the dashed line is the 
curve for $\psi_{30}$, while the  dotted line is $\psi_{20}$. 
These wave functions are dimensionless due to the manner in which they 
are  normalized.}
\label{arwavefunctions}
\end{minipage}
\end{figure}

The inelasticities for the data set LI are shown in 
Fig.~\ref{arinel}, and the phase shifts are shown in
Fig.~\ref{arfullphase}.  The presence of the low mass ``Roper''
resonance can be seen in the  kink in the LI data set at about 500
MeV, which is at the onset of inelasticity. 
Fig.~\ref{arwavefunctions} shows the deuteron  wave  function for the
LI data set and corresponding model.  It is clear that the inelastic 
effects are much stronger than in the previous case; the $\psi _{2}$
and  $\psi _{3}$ components are about ten times stronger than
before.

%
%
\begin{table}[h]
\caption{Binding energy of the deuteron as calculated in each of the 
large inelastic cases.  The percent error shown is the error between 
the calculated value and the expected value 2.22 MeV.  The energies 
are given in MeV.}
\label{armassdeut}
\begin{tabular}{ddd}
 & LI Data & Model 1LI \\
\tableline
Binding Energy & 2.47 & 2.33 \\
Percent Error & 11.3 & 5.0 \\
\end{tabular}
\end{table}

%

%

%

\begin{figure}[t]
\begin{center}
\mbox{
   \epsfxsize=3.5in
\epsfbox{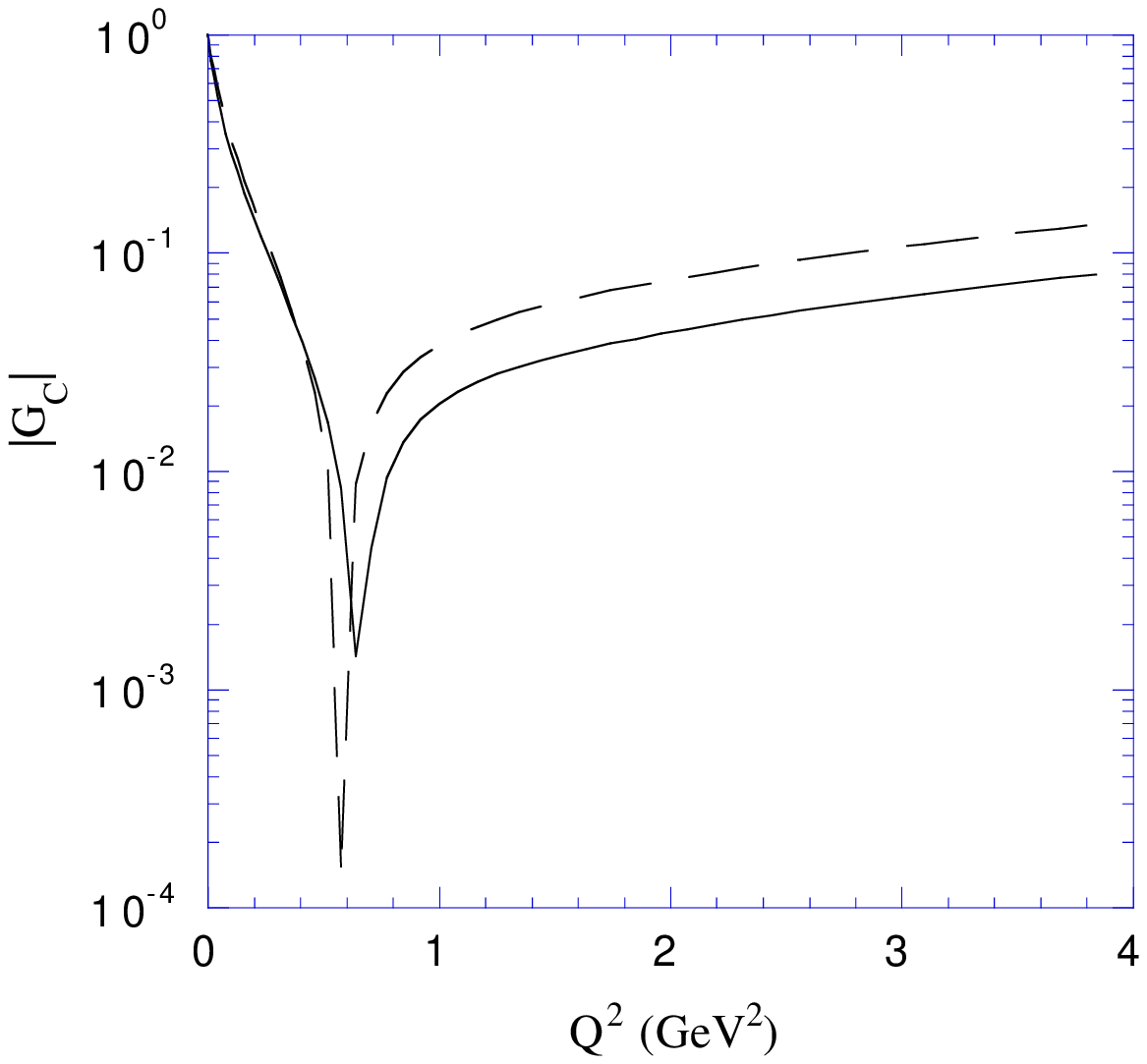}
}
\end{center}
\caption{The deuteron form factor for the large inelastic model and 
data set.   The curves are as in 
Fig.~11.}
\label{ardff}
\end{figure}

Finally, we turn to the most noticable demonstration of the effects of
the  resonances: the deuteron form factor in the large inelastic
case, shown in Fig.~\ref{ardff}. Note the pronounced splitting
between the model and the data, as well as the slight shift in the
zero.  Clearly, the resonances are finally having an effect.  Our
thoughts on this matter will be summarized in the next section.

%

%
%
\subsection{Conclusions}
\label{conclusions}

The major conclusions of this paper are summarized as follows.

\begin{enumerate}
\item The low energy $NN$ data can be fit by tower models using either the proton alone (Model 1)
or a tower consisting of the proton, the Roper, and the ${\rm
D}_{13}$ (Models 3 and 3I).   Model 1 has three parameters in both
the singlet and triplet case;  Models 3 and 3I have five parameters
each.  It has been noted that the fits do not depend strongly on the
coupling constants, but rather on the three form factor parameters. 
Therefore, certain symmetries have been exploited in fixing several
of the remaining coupling constants.  In the case of Model 3I, the
Roper mass is set equal to 1.17 GeV in order to ensure that some
inelasticity appears within the range of interest ($T_{{\rm LAB}} =
0-1.2$ GeV).  Only slight effects are seen from this inelasticity on
such properties of the deuteron and its wave function and form
factor.  
\item Three-member tower models which have small inelasticity fit the 
data just as well as models having no inelasticity.  The differences
between these models and a model with no higher order resonances are
slight, and thus one might conclude that no further investigation is
warranted.  However, unless models with large inelasticity are
explored, we cannot be sure whether the effects are small because the
inelasticities themselves are small, or because inelasticity
truly has no effect on the system whatsoever.
\item In order to explore the effects of large inelasticity, an
adjusted data set was produced by increasing the Roper couplings in
Model 3I.  This large inelasticity data set was then fit below 350
MeV using a proton-only tower model.  Much larger effects of the
resonances are seen in this case.  In this case of unrealistically
large inelasticity the deuteron form factor is changed at
large $Q^2$ by about a factor of 2, which can certainly be detected
by the precision experments which will be carried out at Jefferson
Lab. 
\end{enumerate}

In summary, our simple model suggests that the impact of nucleon
resonances on the deuteron form factor will be substantial
only if the inelasticity in the deuteron channel is substantial
(which is not the case in the $S$-wave channel).  However, we have not
explored the effects of spin, the coupling to the $D$-wave channel, or
the excitation of nucleon resonances by the electromagnetic current,
so our conclusions can only be very preliminary.  Further study is
certainly warranted to assess the sizes of these additional effects.

This concludes our overview and summary of this 
work.  The following sections will discuss the work in greater detail,
beginning with the definition of the general tower of states model,
and concluding with model-specific results.  

%
%
%
%
\section{General theory}
\label{general}

In this part the general framework of the tower of states model will 
be presented.  The relativistic scattering 
equation is obtained, and an analytic solution for the $M$-matrix 
produced. Locations of bound state poles are also discussed.  The 
subject of unitarity is then discussed and relations for the phase 
shifts obtained.  Next, the inelastic  scattering matrix is developed
and the relations for the phase shift and the inelasticity parameter
derived.  Then the gauge invariance of the model is explored. 
Finally, we  derive the expression  for the deuteron form factor.

%
%
\subsection{General description of the tower of states model}
\label{generalform}

We define the tower of states to be the proton and all 
of its charged excited states, and assume that the neutron remains
unexcited by the interaction while the proton can be transformed
into any of its excited states.  The interaction which carries the
$i$th member of this tower (and a neutron) into the $j$th member
(and a neutron) is illustrated in Fig.~\ref{kernel} and takes the
form
\begin{equation}
{\cal V}_{ij} = g_{ij}f\left[x(p_{n}^{\prime},D)\right]
f\left[x(p_{n},D)\right]\, ,
\end{equation} 
where $x(p_{n},D)$ is defined by
\begin{equation}
x(p_{n},D) = p_{n}^{2} - \frac{(p_{n} \cdot D)^2}{D^{2}}\,  ,  
\label{ex}
\end{equation}
with $p_n$ the four-momentum of the neutron and $D$ the total
four-momentum.  Also, $g_{ij}$ is the strong coupling between
the $i$th and $j$th tower members, and $f$ is the Tabakin-style form
factor defined in Eq.~(\ref{formfactor}).  For simplicity we
will sometimes use the notation $x = x(p_{n},D)$ and $x^{\prime} =
x(p_{n}^{\prime},D)$;  when necessary the arguments will be written
in full to avoid confusion.  The variable $x$ is
unchanged by the substitution $p_n\to p_n+\lambda D$, so that
$x(p_{n},D)= x(p,D)$.  Note also that our form factor is {\em not}
equal to unity when both  particles are on the mass shell, as is the
case with many form factors.  

It is important to remember that the neutron is placed consistently on
the mass shell at all times and is {\em not} a member of the tower
of states.  The  proton and neutron are treated as non-identical
particles, which is possible if we limit ourselves to the $np$
system only (which includes the deuteron). The coupling of the photon
to the neutron will be neglected in this simple, begining model.  The
subscript {\em o} shall be used to indicate quantities associated
with the  neutron, while quantities associated with the $i$th
tower member, such as masses, will be given a subscript $i$. 
However, the four-momentum of the neutron will be labeled with 
$n$ to avoid confusion with the $0$th component of the vector $p$,
while the tower particle four-momentum will be written without
subscripts.  The quantities $m_{1} = m_{o} = 0.93825$ GeV will be
used  interchangeably. Outgoing momenta  are distinguished from
incoming momenta by the use of the prime, and internal momenta are
generally denoted by {\em k}, without subscripts.  Finally,
quantities written in {\bf bold face} are three-vectors.

%
%
%
\subsection{Analytic solution for the $M$-matrix and bound state wave 
functions}
\label{analytic}

We would like to find an analytic form for the scattering matrix 
\cite{g93book}, to simplify  the calculation of phase shifts and 
other such important quantities.  Begin with the most general
form for a two-body scattering equation
\begin{equation}
M(p,p^{\prime};D) = {\cal V}(p,p^{\prime};D) + \int_{k}{\cal V}(p,k;D)
G(k,D)M(k,p^{\prime};D)\, ,
\label{gentbs}
\end{equation}
where {\em M} is the scattering amplitude, ${\cal V}$ is the kernel, 
$\int_{k}$  represents an integration over the unspecified internal
four-momentum, and {\em G} is the  two-body propagator.  The exact
forms of these terms will be discussed in  Sec.~\ref{tower}.

\begin{figure}[t]
\begin{center}
\mbox{
   \epsfxsize=4.0in
\epsfbox{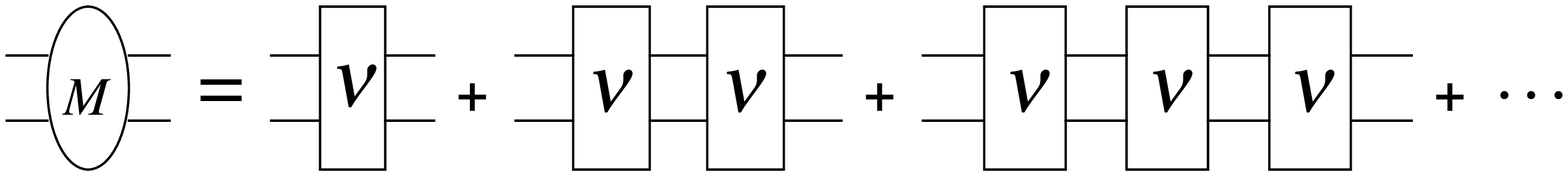}
}
\end{center}
\caption{Diagrammatic form of Eq.~(2.4).}
\label{geomserpic}
\end{figure}

Assuming that the kernel is sufficiently small, perturbation theory 
permits us to obtain an iterative solution to the above equation.  
This solution takes the form
\widetext
\begin{equation}
M = {\cal V} + \int{\cal V} G{\cal V} + \int\int{\cal V} G{\cal V} G
{\cal V} + \cdots +\left(\int{\cal V} G\right)^{n} {\cal V} + \cdots
\, ,
\label{geomseries}
\end{equation}
\narrowtext
also shown diagrammatically in Fig.~\ref{geomserpic}.  This may be 
recognized as a geometric series, which has a simple analytic form,
\begin{equation}
M = \left(1 - \int{\cal V} G\right)^{-1}{\cal V}\, .
\label{geomsum}
\end{equation}
Note that because of the form of the kernel, each term in the sum 
(\ref{geomseries}) will contain a factor of $f(x^{\prime})f(x)$, one 
form factor each for the incoming and outgoing tower member.  
It therefore becomes convenient to define the reduced $M$-matrix  
${\cal M}$,  such that its elements are defined by $M_{ij} =
f(x^{\prime})f(x){\cal M}_{ij}$.

%

In terms of matrices, Eq.~(\ref{geomsum}) can be written 
\begin{equation}
{\cal M} = (1-g{\cal G})^{-1}g\, ,
\label{squigmelems}
\end{equation}
where the matrix ${\cal G}$ is a diagonal matrix of propagators
with elements
\begin{equation}
{\cal G}_{\ell} = \int_{k}f^{2}(k)G_{\ell}\, ,
\label{intgdef}
\end{equation}
and $g$ is the matrix of the strong coupling  constants $\{g_{ij}\}$
defined in Sec.~\ref{generalform} and $G_{\ell}$ is the propagator
for the $\ell$th tower member.  

To locate the bound state poles (if any) of
Eq.~(\ref{geomsum}) introduce the matrix $A = (1-g{\cal G})$ and
write the solution (\ref{squigmelems}) as
\begin{equation}
{\cal M} = \frac{C^{\rm T}g}{\det A}\, ,
\label{redmeq}
\end{equation}
where $C$ is the cofactor matrix of {\it A}.  The full
matrix $M$ is then
\begin{equation}
M = f(x^{\prime})\frac{C^{\rm T}g}{\det A}f(x)\, .
\label{fullm}
\end{equation}
Written in this form it is clear that the locations of any bound state
poles can be found simply  by solving the equation $\det A = 0$.  

To find the bound state wave function, recall that the integral 
equation for the bound state is
\begin{equation}
\Gamma (p,D) = \int_{k}{\cal V} (p,k;D)G(k,D)\Gamma (k,D)\,  ,
\label{bsinteq}
\end{equation}
where $\Gamma$ is the bound state vertex function.
This equation can also be written in matrix notation 
\begin{equation}
(1-{\cal VG})\Gamma = 0\, .
\label{bsvmatrixeq}
\end{equation}
Since the interaction is separable, the solution to this equation
is   
\begin{equation}
\Gamma (p,D) = \left(\begin{array}{c}
c_{1} \\
c_{2} \\
\vdots \\
\end{array}\right) f(x)
\label{bsvertex}
\end{equation}
for as many tower members as we choose.  The bound state wave
function is
\begin{equation}
\phi = {\cal N}G \Gamma,
\label{wfeq}
\end{equation}
where ${\cal N}$ is a normalization constant and $G$ is the diagonal
matrix of propagators with elements $G_{\ell}$.  Therefore the
solution of Eq.~(\ref{bsvmatrixeq}) determines the wave function up
to a normalization constant.

%
%
%
\subsection{Unitarity}
\label{unitarity}

The most general unitarity relation, derivable from
Eq.~(\ref{gentbs}) \cite{g93book}, is
\begin{eqnarray}
M - \bar{M} =&& \int_{k}\bar{M}(G - \bar{G})M \nonumber\\
=&&M^{\ast}({\cal G} - {\cal G}^{\ast})M \, ,
\label{genunit}
\end{eqnarray}
where the last is the matrix version of the unitary relation.  Note
that the bars may be changed to complex conjugates since there is no
Dirac structure in the problem, and hence the equation reduces to
the following conditions on the matrix elements of $M$
\begin{equation}
2i \,{\rm Im}M_{ij} = M_{i\ell}^{\ast}({\cal G}_{\ell} -  {\cal
G}_{\ell}^{\ast})M_{\ell j}\, ,
\label{indunitrel}
\end{equation}
where the implied summation over $\ell$ runs over all members of the 
tower. 

If we are below the production threshold for any tower member except 
the proton,
\begin{equation}
\rho _{\ell} = \frac{{\cal G}_{\ell}-{\cal G}_{\ell}^{\ast}}{2i} = 0 
\quad  \ell > 1\, .
\label{rhol}
\end{equation}
This allows us to simplify Eq.~(\ref{indunitrel}) to
\begin{equation}
{\rm Im}M_{ij} = \rho _{1}M_{i1}^{\ast}M_{1j}.
\label{genimmij}
\end{equation}
For determining the phase shifts, only
the expression for $M_{11}$ need be considered, and
\begin{equation}
{\rm Im}M_{11} = \rho _{1}|M_{11}|^{2}\, ,
\label{immoo}
\end{equation}
where $\rho_{1}$ is the relativisitic phase space factor
\begin{equation}
\rho_{1} = \frac{\sqrt{W^{2}-4m_{o}^{2}}}{8\pi W}\, .
\label{phsspc}
\end{equation}
Note the similarity between Eq.~(\ref{immoo}) and the unitarity 
relation for the $\ell$th partial wave in nonrelativistic
scattering
\begin{equation}
{\rm Im}f_{\ell} = \frac{p}{K}|f_{\ell}|^{2},
\label{nrf}
\end{equation}
where {\em K} is a constant, usually equal to unity in 
nonrelativistic theory. Therefore, the standard identifications may
be made,
\begin{eqnarray}
f_{\ell} & \leftrightarrow & -M_{11}  \nonumber\\
\frac{p}{K} & \leftrightarrow & \rho _{1}\, , \label{rnrid}
\end{eqnarray}
(where the minus sign in the first relation comes from a comparison
of the integral equations and not the unitary relation) allowing us 
to write
\begin{equation}
M_{11} = -\frac{e^{i\delta _{1}}\sin \delta _{1}}{\rho _{1}}\, ,
\label{moo}
\end{equation}
which implies
\begin{eqnarray}
{\rm Re}M_{11} & = & -\frac{\cos \delta _{1}\sin \delta _{1}} {\rho
_{1}}  \nonumber\\ 
{\rm Im}M_{11} & = & -\frac{\sin ^{2}\delta
_{1}}{\rho _{1}} \, .\label{imandre}  
\end{eqnarray}
These can be combined to yield a single expression for $\delta _{1}$,
\begin{equation}
\delta _{1} = \arctan \left( \frac{{\rm Im}M_{11}}{{\rm Re}M_{11}}
\right).
\label{delta}
\end{equation}
This is the general expression for the elastic phase shift.  In order 
to fit to the phase  shift data, the real and imaginary parts of
$M_{11}$ must be written in terms of the free parameters.  This will
be discussed in Sec.~\ref{tower}.

%
%
%
\subsection{Inelasticity}
\label{inelasticity}

To explore the inelastic scattering region for the tower model
\cite{gs93} we generalize the structure of the scattering matrix
to include the possibility of inelasticity:
\begin{equation}
-M_{11} = \frac{\eta e^{2i\delta_{1}} - 1}{2i\rho_{1}},
\label{inelmatrix}
\end{equation}
where $\eta= \cos\rho<1$ is the inelasticity parameter.  Note that
elastic scattering corresponds to $\eta=1$ (or $\rho=0$), and that
as $\eta$ decreases toward zero (or $\rho$ toward $\pi/2$) the
scattering becomes increasingly inelastic.    

It follows from (\ref{inelmatrix}) that the real and imaginary parts
of $M_{11}$ are now
\begin{eqnarray}
{\rm Im}M_{I} & = & \frac{1}{\rho_{1}}(\eta \cos 2\delta _{1} -1) 
\nonumber \\ {\rm Re}M_{I} & = &
-\frac{1}{\rho_{1}}\eta \sin 2\delta _{1}\, . \label{inelimre} 
\end{eqnarray}
Solving for the two unknowns, $\eta$ and
$\delta _{1}$, gives
\begin{eqnarray}
\eta & = & \frac{\rho_{1}{\rm Im}M_{11} + 1}{\cos 2\delta _{1}} 
\nonumber\\
\tan 2\delta _{1} & = & -\frac{{\rm Re}M_{11}}{{\rm Im}M_{11} + 
{\displaystyle\frac{1}{\rho_{1}}}}\, . \label{etandelta} 
\end{eqnarray}
We will return to these expressions for the phase shift and
inelasticity parameter in Sec.~\ref{tower}.

%
%
%
\subsection{Electromagnetic current and gauge invariance}
\label{invariance}

In this section we describe the one and two body currents, and show 
that  the model is exactly gauge invariant. 

Charge conservation requires that the coupling between the photon and
a given tower member be of the form $e_{ij} =\zeta_{ij}e$, where 
{\it e} is the electromagnetic charge of the proton and
$\zeta_{ij}$ is a scaling factor which must equal unity if $i=j$.
The off-diagonal terms are not constrained by charge conservation,
but the simplest way to insure that the model is gauge invarinat is
to require them to be be transverse, that is, $q_{\mu}j^{\mu} = 0$. 
Finally, we require that the diagonal component of the current
satisfy a Ward-Takahashi identity \cite{t57},\cite{w50nc}, which may
be expressed in a compact fashion,
\begin{equation}
q_{\mu}j^{\mu}_{ij}(p',p) = -e\left[G_{i}^{-1}\left(p^{\prime
2}\right) - G_{j}^{-1}\left(p^{2}\right)\right]\delta_{ij} \, ,
\label{wardtakaid}
\end{equation}
where $p$ and $p'$ are the initial and final momentum of the tower
member, respectively.   This restriction is satisfied by 
the following form of the electromagnetic current of the
tower member:
\begin{equation}
j_{ij}^{\mu}(p',p) = e_{ij}F_{ij}(Q^{2})\left( P^{\mu} -
\frac{P\cdot q} {q^{2}}q^{\mu}\right) +
e_{ij}\delta_{ij}\left(\frac{P\cdot q}{q^{2}}q^{\mu}\right)\, ,
\label{currents}
\end{equation}
where $P = p^{\prime} + p$ and the form factors $F_{ij}(Q^2)$
represent  the structure of the current due to the true
(nonpointlike) nature of  the tower members.  To avoid kinematic
singularities we require 
\begin{eqnarray}
F_{ii}(0) =&& 1\nonumber\\
F_{ij}(0) =&& 0 \, .
\end{eqnarray}
Except for these requirements, the form factors need not be further
specified at this time.

Because the strong form factors $f$ depend on the total
four-momentum, we expect them to generate an interaction current. 
Using a technique similar to  that of Ito, Buck and Gross
\cite{ibg91}, and assuming that $\zeta_{ij}=\delta_{ij}$ for
simplicity, leads to the following interaction current
(see the Appendix for details)
\widetext
\begin{eqnarray}
j_{ij}^{\mu^{INT}}(p_{n}^{\prime}, p_{n};D^{\prime},D) = 
e g_{i j}\Biggl\{&& {\cal Z}^{\prime
\mu}\frac{f\left[x(p_{n}^{\prime}, D^{\prime})\right] -
f\left[x(p_{n}^{\prime}, D)\right]}{x(p_{n}^{\prime}, D^{\prime}) -
x(p_{n}^{\prime}, D)}f\left[x(p_{n},
D)\right]\nonumber\\
 && + {\cal Z}^{\mu}f\left[x(p_{n}^{\prime}, 
D^{\prime})\right]\frac{f\left[x(p_{n}, D^{\prime})\right] -
f\left[x(p_{n}, D)\right]}{x(p_{n}, D^{\prime}) - x(p_{n},
D)}\Biggr\} \, , \label{intcurr}
\end{eqnarray}
where
\begin{eqnarray}
{\cal Z}^{\mu} = \frac{1}{2D^{2}D^{\prime 2}}\biggl\{ && (D +
D^{\prime})^{\mu}\left[(p_{n}\cdot D)^{2} + (p_{n}\cdot
D^{\prime})^{2}\right]
\nonumber\\ 
&&-p_{n}^{\mu}\;\;\;\; p_{n}\cdot\! (D + D^{\prime}) \left(D^2 + 
D'^2 \right) \biggr\}  \,  , \label{zee} 
\end{eqnarray}
and ${\cal Z}'^{\mu}$ is obtained from ${\cal Z}^{\mu}$ by 
replacing
$p_n$ by $p'_n$.  Contracting $q_{\mu}$ into this current gives
\begin{eqnarray}
q_{\mu}j_{ij}^{\mu^{INT}}(p_{n}^{\prime},
p_{n};D^{\prime},D)  =  e \biggl\{{\cal V}_{i
j}(p_{n}^{\prime}, p_{n};D') - {\cal V}_{i j}(p_{n}^{\prime},
p_{n};D) \biggr\} \, .\label{qincurr}
\end{eqnarray}
\narrowtext

\begin{figure}[t]
\begin{center}
\mbox{
   \epsfxsize=4.5in
\epsfbox{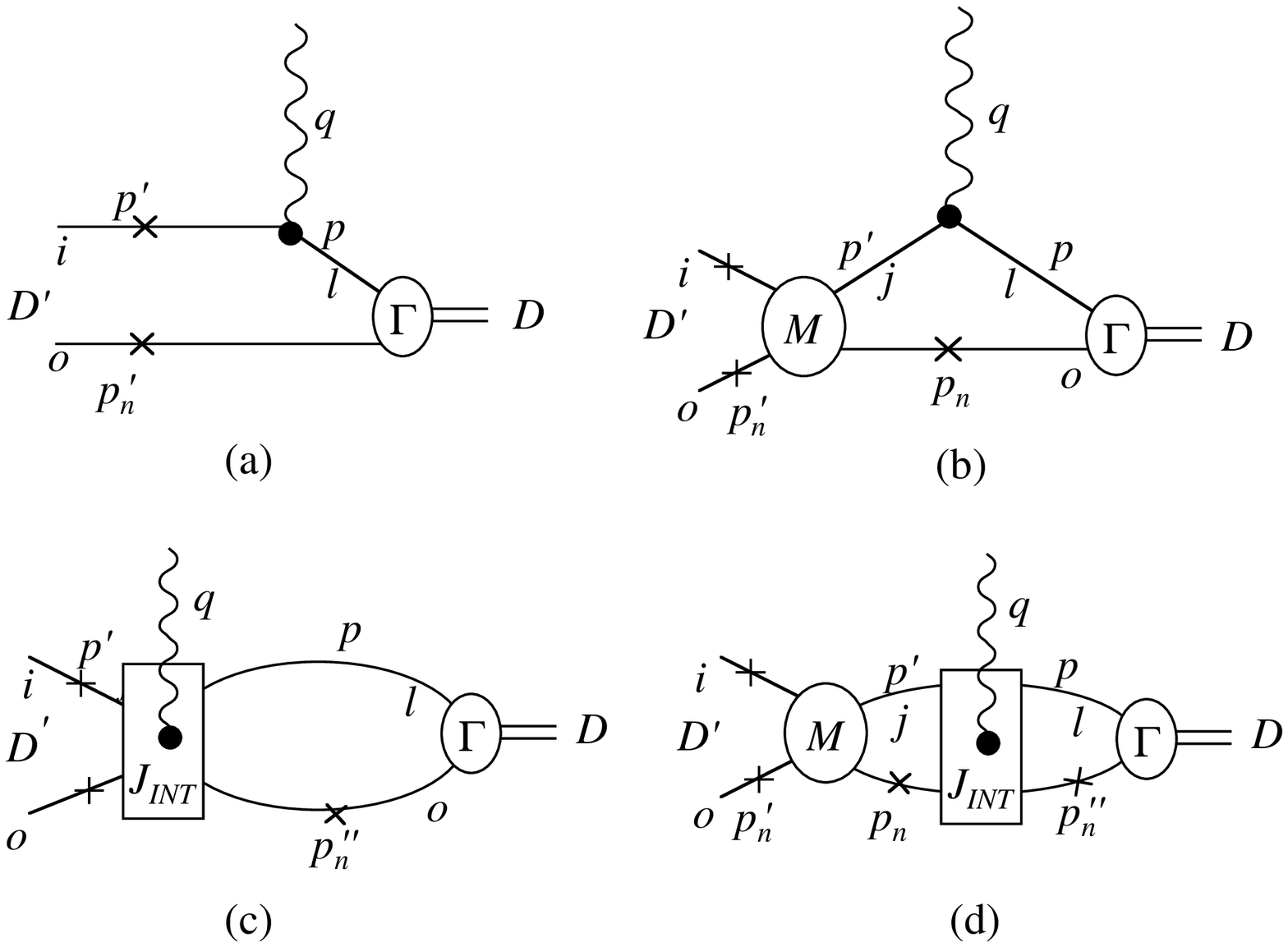}
}
\end{center}
\caption{The four diagrams required for the calculation of the
electrodisintegration process.  The large boxes with photons coupling
to them are interaction currents, the oval with a $\Gamma$
is a bound state vertex, and the circle with an $M$
is a scattering matrix.}
\label{gidiags}
\end{figure}

Recalling that the bound state is generated by the infinte sum
given in Eq.~(\ref{geomseries}), it can be shown that the
electrodisintegration process is described by the four diagrams 
shown in  Fig.~\ref{gidiags}.  We will show that these diagrams,
together with the conditions (\ref{wardtakaid}) and
(\ref{qincurr}), are sufficient to ensure that the
electrodisintegration process is gauge invariant. 

%

Note first that all four diagrams have the same incoming state, 
namely
\begin{equation}
\phi_{\ell}^{i} = \phi_{\ell}^{B}(p_n,D) = G_{\ell}(p) 
\Gamma_{\ell}(p_n,D),  \label{inphi}
\end{equation}
where $\phi_{\ell}^{B}$ is the bound state wave function of a 
neutron and the $\ell$th member of the tower of states.  The final
states, however,  are  different.  For diagrams \ref{gidiags}(a) 
and \ref{gidiags}(c) there is no final state interaction and the
final state is a plane wave state with
\begin{equation}
\bar{\phi}_{ij\;{\rm pw}}^{f}(p'_n,p_n;D') =2E \,\delta_{ij}\,
(2\pi)^3 \delta^{3}(p'_{n}-p_{n}) \, ,
\label{finphiac}
\end{equation}
where $E=\sqrt{m_n^2 + p_n^2}$.
Diagrams \ref{gidiags}(b) and \ref{gidiags}(d) include the final 
state interactions.  In the notation of the figures, this is
\begin{equation}
\bar{\phi}_{ij\;{\rm scatt}}^{f}(p'_n,p_n;D') =
\bar{M}_{ij}(p_{n}^{\prime},p_{n};D^{\prime}) G_{j}(p').
\label{finphidb}
\end{equation}
These two processes can be combined by introducing the total final
state wave function
\widetext
\begin{eqnarray}
\bar{\phi}_{ij}^{f}(p'_n,p_n;D') & = & \bar{\phi}_{ij\;{\rm
pw}}^{f}(p'_n,p_n;D') + \bar{\phi}_{ij\;{\rm
scatt}}^{f}(p'_n,p_n;D') 
\nonumber\\
&=&\bar{M}_{ij}(p_{n}^{\prime},p_{n};D') G_{j}(p')
+2E\,\delta_{ij}\,(2\pi)^3\delta^{3}(p_{n}^{\prime}-p_{n}) 
\,  .\nonumber
\label{combphis}
\end{eqnarray}
Using this notation, the four diagrams shown in
Fig.~\ref{gidiags} can be written
\begin{eqnarray}
\langle J^{\mu}\rangle_{i} =&& \int {d^{3}p_n\over
(2\pi)^3 2 E} \;\bar{\phi}_{ij}^{f}(p'_n,p_n;D')
j_{j\ell}^{\mu}(p',p)\phi_{\ell}^{B}(p_n,D)\nonumber\\
&&+
\int\frac{d^{3}p_n\;d^{3}p^{\prime\prime}_n}{(2\pi)^{6} 4 E E''}\,
\bar{\phi}_{ij}^{f}(p'_n,p_n;D')\,
j_{j\ell}^{\mu^{INT}}(p,p^{\prime\prime};D^{\prime},D)\,
\phi_{\ell}^{B}(p''_n,D)\, ,
\label{expecval}
\end{eqnarray}
where summation over repeated indicies is implied.

Now, to demonstrate that the current (\ref{expecval}) is gauge
invariant we contract it with
$q_{\mu}$ and use the relations (\ref{wardtakaid}) and
(\ref{qincurr}) to obtain
\begin{eqnarray}
q_{\mu}\langle J^{\mu}\rangle_{i} =& & -e \int {d^{3}p_n \over
(2\pi)^3 2E} \bar{\phi}_{ij}^{f}(p'_n,p_n;D')\left\{G_{j}^{-1}
\left(p^{\prime}\right) - G_{j}^{-1}(p)\right\}\phi_{j}^{B}(p_n,D)
\nonumber\\
& & +e\int\frac{d^{3}p_n\;d^{3}p''_n}{\left(2\pi\right)^{6}4EE''}
\;\bar{\phi}_{ij}^{f}(p'_n,p_n;D') \nonumber\\
&&\quad\times\biggl\{ {\cal
V}_{j\ell}(p_{n},p_{n}'';D') -
{\cal V}_{j\ell}(p_{n},p_{n}'';D)\biggr\}
\phi_{\ell}^{B}(p''_n,D) \label{qinexpecj} \, .
\end{eqnarray}
We can reduce the expressions $G_{j}^{-1}(p)\phi_{j}^{B}$
and $\bar{\phi}_{ij}^{f}G_{j}^{-1}(p')$ which occur in the 
first  term of Eq.~(\ref{qinexpecj}) by using the fact that 
$\bar{\phi}_{ij}^{f}$ and $\phi_{j}^{B}$ satisfy the spectator 
equation:
\begin{equation}
G_{j}^{-1}(p)\phi_{j}^{B}(p_n,D) = \int\frac{d^{3}p''_n}
{(2\pi)^{3} 2 E''}{{\cal
V}}_{jm}(p_{n},p_{n}'';D)\phi_{m}^{B}(p''_n,D)\, ,
\label{speceqn}
\end{equation}
and similiarly for the final state. 
Using these equations to rewrite the first term of
Eq.~(\ref{qinexpecj}) yields
\begin{eqnarray}
q_{\mu}\langle J^{\mu}\rangle_{i} =&&- e\int\frac{d^{3}p_n\;d^{3}
p''_n}{\left(2\pi\right)^{6} 4 EE''}
\biggl\{\bar{\phi}_{im}^{f}(p'_n,p''_n;D') {\cal
V}_{mj}\left(p''_{n},p_{n};D^{\prime}\right)
\phi_{j}^{B}(p_n,D)
\nonumber\\
&&\qquad\qquad\qquad\qquad-\bar{\phi}_{ij}^{f}(p'_n,p_n;D') 
{{\cal V}}_{jm}\left(p_{n},p''_{n};D\right)
\phi_{m}^{B}(p''_n,D)\biggr\}
\nonumber\\
 && +e\int\frac{d^{3}p_n\;d^{3}p''_n}{\left(2\pi\right)^{6}4EE''}
\;\bar{\phi}_{ij}^{f}(p'_n,p_n;D') \nonumber\\
&&\quad\times\biggl\{ {\cal
V}_{j\ell}(p_{n},p_{n}'';D') -
{\cal V}_{j\ell}(p_{n},p_{n}'';D)\biggr\}
\phi_{\ell}^{B}(p''_n,D)  \nonumber\\
=&& 0\,  , \label{fingiexp}
\end{eqnarray}
where the cancellation follows from interchanging the $p_n$
and $p''_n$ integrations in the first term and relabeling the
sums.  This completes our proof that the current is gauge
invariant.
%
%
%
\subsection{Deuteron form factor}
\label{deuteron}

%
\begin{figure}
\vspace*{-3in}
\begin{center}
\mbox{
   \epsfxsize=3.5in
\epsfbox{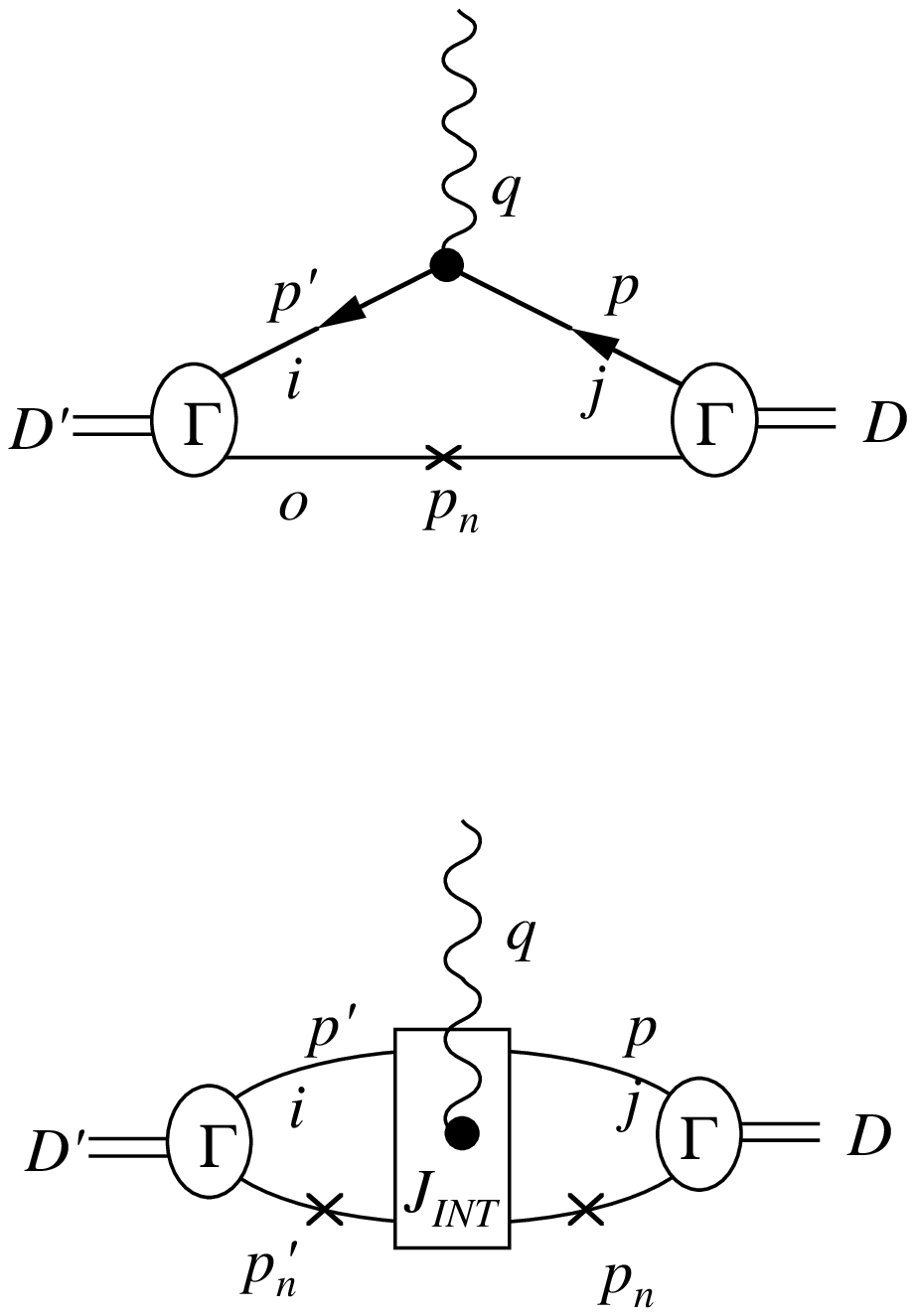}
}
\end{center}
\caption{The diagrams needed to calculate the deuteron form factor. 
Diagram (a) is the relativistic impulse approximation
and diagram (b) includes the interaction current.}
\label{dfftsdiag}
\end{figure}

In this section we discuss the calculation of the deuteron current
given by the two diagrams shown in Fig.~\ref{dfftsdiag}.  The
deuteron form factor, $G_C$, is defined by the relation
\begin{equation}
J_{p}^\mu = e\,G_C(Q^2)\,(D + D^{\prime})^{\mu}\,  .
\label{pointm}
\end{equation}
The form factor describes the structure of the deuteron; the current
for a structureless point particle is obtained from this equation
by replacing $G_C \to 1$.
  
The current for the processes shown in Fig.~\ref{dfftsdiag} is
\begin{eqnarray}
J_{D}^\mu =&&
\int\frac{d^{3}p_{n}}{(2\pi)^{3} 2E}
\;\bar{\phi}_{i}^{B}(p_n,D') j_{ij}^{\mu}(p',p)\phi_{j}^{B}(p_n,D)
\nonumber\\
&&+\int\frac{d^{3}p_{n}^{\prime}\,d^{3}p_{n}}{(2\pi)^{6}4E E'}
\;\bar{\phi}_{i}^{B}(p'_n,D')j^{\mu^{INT}}_{ij}(p_n',p_n;D',D)
\phi_{j}^{B}(p_n,D)\, ,\label{mt}
\end{eqnarray}
\narrowtext
where the one-body and interaction currents are the same as those
given in Eqs.~(\ref{currents}) and (\ref{intcurr}).  When the
polarization vectors $\epsilon _{\mu}$ of the photon (including
the longitudinal polarization vector associated with a virtual
photon) are  contracted into these currents, any terms
containing $q^{\mu}$ will vanish.  Using the notation
shown in the relativistic impulse approximation (RIA) diagram
given in Fig.~\ref{dfftsdiag} leads to the replacement $P^{\mu} = 
(D + D^{\prime})^{\mu} - 2p_{n}^{\mu}$,  and the
effective  one-body current is simply
\begin{equation}
j_{ij}^{\mu} = e F(Q^{2})\biggl\{ (D + D^{\prime})^{\mu} -
2p_{n}^{\mu}\biggr\}\,\delta_{ij}\, .
\label{effj}
\end{equation}
Since the interaction current does not contain any factors of 
$q_{\mu}$, it remains as given in Eq.~(\ref{intcurr}).

We now extract an explicit formula for the deuteron form factor
from Eq.~(\ref{mt}).  First, note that the integrand of the
integral 
\begin{equation}
I_1=\int\frac{d^{3}p_{n}}{(2\pi)^{3} 2E}
\;\bar{\phi}_{i}^{B}(p_n,D') \;\phi_{i}^{B}(p_n,D) \, ,
\end{equation}
is an even function of the three vector ${\bf p}_n$ in the Breit
frame, and hence in this frame 
\widetext
\begin{eqnarray}
I^\mu_p = &&\int \frac{d^{3}p_{n}}{(2\pi)^{3} 2E }
\;\bar{\phi}_{i}^{B} (p_n,D') \;p_{n}^{\mu}\;\phi_{i}^{B}(p_n,D)
\nonumber\\
= &&(D'+D)^\mu \int \frac{d^{3}p_{n}}{(2\pi)^{3} 2E }
\;\bar{\phi}_{i}^{B} (p_n,D') \;\left[\frac{p_{n}\cdot (D +
D^{\prime})} {(D + D^{\prime})^{2}}\right]\;\phi_{i}^{B}(p_n,D)\, .
\label{pint}
\end{eqnarray}
Since both sides of this equation are covariant,  this expression
holds in any frame.  

The interaction current term can also be simplified in the same
way.  Since the initial and final states are identical, the
integrand of the interaction integral is symmetric under the
interchange ${\bf p}_n\leftrightarrow -{\bf p}'_n$ in the Breit
frame [which transforms $x(p'_n,D')$ into $x(p_n,D)$ and
$x(p'_n,D)$ into $x(p_n,D')$], {\it except\/} for the terms in 
${\cal Z}^\mu$ and ${\cal Z}'^\mu$  which are linear in the
three-vector parts of $p_n^\mu$ and $p'^\mu_n$.  These terms are
odd under this interchange, and thus vanish, leaving only the time
components of $p_n^\mu$ and $p'^\mu_n$.  The interaction current
can therefore be written as the product of two integrals
\begin{eqnarray}
J_{D\,{\rm INT}}^\mu = 2e g_{ij}\; &&\int\frac{d^{3}p'_{n}}
{(2\pi)^{3}2 E'}
\;{\phi}_{i}^{B}(p'_n,D') f\left[x(p_{n}', D^{\prime})\right]
\nonumber\\
\times&&\int\frac{d^{3}p_{n}}{(2\pi)^{3}2 E} 
\phi_{j}^{B}(p_n,D)
{\cal Z}^{\mu}\left(\frac{f\left[x(p_{n}, D^{\prime})\right] -
f\left[x(p_{n}, D)\right]}{x(p_{n}, D^{\prime}) - x(p_{n},
D)}\right)\, , \label{int1}
\end{eqnarray}
where, for the deuteron form factor with $D^2=D'^2=M_d^2$, ${\cal
Z}^\mu$ simplifies to
\begin{eqnarray}
{\cal Z}^\mu =&& (D + D^{\prime})^{\mu}\;
\overline{{\cal Z}}\nonumber\\  
=&&(D +D^{\prime})^{\mu}\;\left\{ {(p_{n}\cdot D)^{2} + (p_{n}\cdot
D^{\prime})^{2}\over 2M_d^4}  -
 {\left[p_n\cdot( D+ D')\right]^2 \over M_d^2\, (D'+D)^2}\right\}
\, . \label{zbar}
\end{eqnarray}
Each of the two integrals in (\ref{int1}) is covariant, and
therefore can be evaluated in any frame.  The first of these two
integrals depends on $D'$ only, and is therefore a constant
(which is obvious when it is evaluated in the rest frame of the
outgoing deuteron).  Denoting this constant by $C_i$
\begin{equation}
C_i=\int\frac{d^{3}p'_{n}}
{(2\pi)^{3}2 E'}
\;{\phi}_{i}^{B}(p'_n,D') f\left[x(p_{n}', D^{\prime})\right]\, ,
\label{ci}
\end{equation}
and using (\ref{pint}) and (\ref{int1}) gives the following
result for the deuteron form factor
\begin{eqnarray}
G_C(Q^2) =&& F(Q^2)\int \frac{d^{3}p_{n}}{(2\pi)^{3} 2E }
\;\bar{\phi}_{i}^{B} (p_n,D') \;\left[1-\frac{2p_{n}\cdot (D +
D^{\prime})} {(D + D^{\prime})^{2}}\right]\;\phi_{i}^{B}(p_n,D)
\nonumber\\
&&+ 2 g_{ij}\; C_i \int\frac{d^{3}p_{n}}{(2\pi)^{3}2 E} 
\phi_{j}^{B}(p_n,D)
\overline{{\cal Z}}\left(\frac{f\left[x(p_{n},
D^{\prime})\right] - f\left[x(p_{n}, D)\right]}{x(p_{n},
D^{\prime}) - x(p_{n}, D)}\right)\, . \label{tdff}
\end{eqnarray}
\narrowtext
Note that the only the first term (the RIA) depends on the nucleon
form factor, $F(Q^2)$.  The second, exchange term, is independent
of this factor and depends only on the strong form factor $f$.  As
we will discuss in the next section, this term actually goes to a
constant as $Q^2\to\infty$.

This concludes the discussion of the general theory of the tower of
states. In the next section we will apply this to the specific
models under consideration.

%
%
%
%
\section{Tower of States Models}
\label{tower}

In this section the detailed form of the equations appropriate for
a three-member tower are presented.  Then the details of the
equations needed to calculate the phase shifts and  inelasticity
parameter for a three-member tower above the inelastic threshold of
the second tower member are presented.  In the final subsection the
behavior of the interaction current contribution to the deuteron
form factor is more carefully examined.

%
%
\subsection{Specific structure for a three-member tower}
\label{specific}

\begin{figure}[t]
\begin{center}
\mbox{
   \epsfxsize=3.0in
\epsfbox{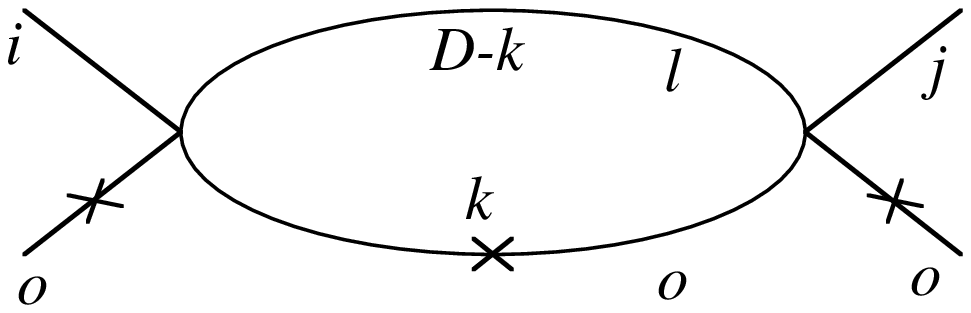}
}
\end{center}
\caption{The definition of the momenta for the bubble diagram.  It is
these  momenta which must be used in the integrals ${\cal G}_{i}$.}
\label{bubble}
\end{figure}

Using the spectator formalism, the generic integrals in
Sec.~\ref{analytic} become
\begin{equation}
\int _{k} = -\int \frac{d^{3}k}{(2\pi )^{3} 2E}\, ,
\label{specint}
\end{equation}
where $E=\sqrt{m_n^2+k^2}$ (where $k^2$ is the square of the {\it
three\/}-momenta in these formulae, not to be confused with the
four-momenta), and the elements of the propagator matrix
(\ref{intgdef}) in the rest frame of the
$NN$ system, where $D=(W,{\bf 0})$, are
\begin{equation}
{\cal G}_{i} = -\int \frac{d^{3}k}{(2\pi)^{3} 2E}\;
\frac{f^{2}[x(k,D)]}{E_{i}^{2} - (W - E)^{2}}\, .
\label{specg}
\end{equation}
The momenta are labeled in Fig.~\ref{bubble}.  Here
$E_i=\sqrt{m_i^2+k^2}$ and, in the rest frame,
\begin{equation}
f\left[x(k,D)\right] = \frac{(\alpha ^{2} + {k}^{2})(p_{c}^{2}
-{k}^{2})} {(\gamma ^{2} +  {k}^{2}) (\beta ^{2} +
{k}^{2})^{2}}\equiv f(k^2)\, .
\label{tvtff}
\end{equation}
%

%

If $W<m_o+m_i$ (where $m_o=m_n$ as discussed above), the integrand
of Eq.~(\ref{specg}) is real for all $k$ and the integral is
real.  When $W>m_o+m_i$ the integrand has a pole in the
range of integration located at ${k} = {k}_{p}^i$, where
\begin{equation}
{k}^i_{p} = \sqrt{\left( \frac{W^{2} - m_{i}^{2} +
m_{o}^{2}}{2W}\right) ^{2} - m_{o}^{2}}\, ,
\label{kpole}
\end{equation}
This pole gives the integral an imaginary part.
The threshold for the onset of this complex behavior, in terms of
the kinetic energy of the nucleon in the $NN$ laboratory system, is
\begin{equation}
T_{{\rm LAB}}^i={(m_o+m_i)^2-4m_o^2\over 2m_o} \, .
\end{equation}
These thresholds are given in Table \ref{thres} for each of the
channels discussed in Sec.~\ref{overview}.

%
\begin{table}[b]
\begin{center}
\begin{minipage}{3.0in}
\caption{Thresholds for the onset of inelasticity for the various
channels.}
\label{thres}
\begin{tabular}{cc}
channel mass & $T_{{\rm LAB}}^i$ \\
in GeV & in MeV \\
\tableline
0.93825 & 0 \\
1.17 & 492 \\
1.44 & 1138 \\
1.52 & 1344 \\
\end{tabular}
\end{minipage}
\end{center}
\end{table}

For energies in the fitting region ($T_{{\rm LAB}} <$ 350
MeV) only ${\cal G}_{1}$ will have an imaginary part. (The case
where more than one of the integrals has an imaginary part will be
discussed in Sec.~\ref{inelasticities}.)   For all energies ${\cal
G}_{1}$ can be written
\begin{equation}
{\cal G}_{1} = {\cal P}\int _{0}^{\infty}{k^2 dk \over
E}\, \frac{f^{2}(k^2)}{W(2E - W)} +
i\,{\rm Im}{\cal G}_{1}\, ,
\label{renimi}
\end{equation}
where ${\cal P}$ denotes the Cauchy Principle Value of the
integral, and the imaginary part is
\begin{equation}
{\rm Im}{\cal G}_{1} = \pi\,\frac{\sqrt{W^{2} - 4m_{o}^{2}}}{4W}
\;f^2(k^2_{1 p})  \, .
\label{imi}
\end{equation}
The real part can be evaluated numerically. 

Now return to Eq.~(\ref{fullm}) and evaluate the $NN$ scattering
amplitude, which in the tower notation is $M_{11}$.  At energies
below the production threshold for either of the resonances, the
imaginary part of $M_{11}$ can come only from ${\cal G}_{1}$,
which contributes only to the denominator $\det (1 - g{\cal G})=
\det A$, and recalling that the cofactor matrix of $A$ is denoted
by $C$, we obtain 
\begin{equation}
M_{11} = \frac{(C^{\rm T}\, g)_{11}}{{\rm Re}(\det A) + i{\rm
Im}(\det A)}\, f(x')f(x)\,  .
\label{expm}
\end{equation}
Rationalizing the denominator gives the real and imaginary parts
of $M_{11}$, and hence the phase shift in the elastic scattering
region depends only on the $\det A$ 
\begin{equation}
\tan \delta _{1} = -\frac{{\rm Im}(\det A)}{{\rm Re}(\det A)}\, .
\label{ttdelta}
\end{equation}
For a three member tower, the real and imaginary parts of $\det
A$ are
\widetext
\begin{eqnarray}
{\rm Im}(\det A) & = & -\left[ g_{11} - (g_{11}g_{33} - g_{13}^{2})
{\cal G}_{3} - (g_{11}g_{22} - g_{12}^{2}){\cal G}_{2} + B_{g}{\cal
G}_{2}{\cal G}_{3}\right] {\rm Im}{\cal G}_{1} \nonumber \\ 
{\rm Re}(\det A) & = & 1 - g_{33}{\cal G}_{3} - g_{22}{\cal G}_{2} 
+ (g_{22}g_{33} - g_{23}^{2}){\cal G}_{2}{\cal G}_{3} \nonumber\\
& & - \left[ g_{11} - (g_{11}g_{33} - g_{13}^{2}){\cal G}_{3} - 
(g_{11}g_{22} - g_{12}^{2}){\cal G}_{2} + B_{g}{\cal G}_{2}{\cal
G}_{3}\right] {\rm Re}{\cal G}_{1} \, , \label{imaandrea} 
\end{eqnarray}
\narrowtext
where $B_{g}$ is 
\begin{equation}
B_{g} = (g_{11}g_{22}g_{33} - g_{11}g_{23}^{2} - g_{12}^{2}g_{33} 
+ 2g_{13}g_{12}g_{23} - g_{13}^{2}g_{22})\, .
\label{bsubg}
\end{equation}

The deuteron vertex function for a three-member tower, 
\begin{equation}
\Gamma  =  \left( \begin{array}{c}
c_{1} \\
c_{2} \\
c_{3} \\
\end{array}\right) f(k^2)\, ,\label{wfmatrices} 
\end{equation}
is obtained from Eq.~(\ref{bsvmatrixeq}). The solution for $c_2$
and $c_3$ in terms of $c_1$ (which must be determined by
normalization) is
\begin{eqnarray}
c_{2} & = & \frac{g_{12}(1-g_{33}{\cal G}_3)  +
g_{13}g_{23}} {(1 - g_{22}{\cal G}_{2})(1-g_{33}{\cal G}_3)-
g_{23}^2 {\cal G}_2}\; {\cal G}_{1} c_1   \nonumber\\  
c_{3} & = & \frac{g_{13}(1-g_{22}{\cal G}_2)  +
g_{12}g_{23}} {(1 - g_{22}{\cal G}_{2})(1-g_{33}{\cal G}_3)-
g_{23}^2 {\cal G}_2}\; {\cal G}_{1} c_1 \, , \label{redcsol}
\end{eqnarray}
where the integrals ${\cal G}_{i}$ must be evaluated at  $W =
M_{D}$.  Normalizing the wave function to unity at $k=0$, the
general expression for the deuteron wave function for a three-member
tower is
\widetext
\begin{equation}
\psi_{i}(k) = \frac{\phi_{i}^{B}({k})}{\phi _{1}^{B}(0)} =
\frac{c_{i}M_D(2m_{o} - M_D)}{c_1 \left(m_{i}^{2} -
m_{o}^{2} + 2M_D\sqrt{m_{o}^{2} + {k}^{2}} - M_D^{2}\right)}\;
{\gamma^{2}\beta^{4} \over\alpha ^{2}p_{c}^{2}} f(k^2)\, .
\label{genttpsi}
\end{equation}
\narrowtext

Finally, we evaluate the deuteron form factor in the case 
of the  three-member tower. The integrations in
Eq.~(\ref{tdff}) are easily carried out if the  momenta in 
Fig.~\ref{dfftsdiag} are evaluated in the Breit frame, where
\begin{eqnarray}
p & = & D - p_{n} \nonumber \\
p^{\prime} & = & D^{\prime} - p_{n}^{\prime} \nonumber \\
q & = & (0, {\bf q}) \label{breitp} \\
D & = & \left( D_0, -{\textstyle\frac{1}{2}}{\bf q}\right) \nonumber \\
D^{\prime} & = & \left( D_0, 
{\textstyle\frac{1}{2}}{\bf q}\right)\, , \nonumber
\end{eqnarray}
with $D_0=\sqrt{M_{D}^{2} + {\bf q}^{2}/{4}}$.  We use the standard
notation $Q^2=-q^2={\bf q}^2$.  For the nucleon form factor, $F(Q^2)$,
we use the dipole approximation,
\begin{equation}
F(Q^{2}) = \left(\frac{1}{1 + 0.71 Q^{2}}\right)^{2}\, .
\label{neff}
\end{equation}
where $Q^2$ is in GeV$^2$.

In the next subsection we extend the results for $M_{11}$ to the
inelastic region.

%
%
%
\subsection{Inelasticities in the three-member tower}
\label{inelasticities}

Since one of the tower models (Model 3I)
has a significant inelastic region,  in this subsection we develop the
relations needed to calculate the phase shifts and inelasticities above
the production threshold for particle 2.  

In Section \ref{specific}, the determination of the real and imaginary
parts of $M$ was considerably simplified by realizing that only terms
containing ${\cal G}_{1}$ could contribute to the imaginary part. 
In the inelastic region ${\cal G}_{2}$ will also
have an imaginary part.  Breaking the 
numerator and the denominator of Eq.~(\ref{expm}) into their real and
imaginary parts and  rationalizing the denominator gives 
\widetext
\begin{eqnarray}
{\rm Im}M_{11} & = & \frac{{\rm Im}(C^{\rm T}g)_{11}{\rm Re}({\det A}) 
- {\rm Re}(C^{\rm }g)_{11}{\rm Im}({\det A})}{\left[{\rm Re}(\det
A)\right]^{2} +
\left[{\rm Im}({\det
A})\right]^{2}}f(x^{\prime})f(x)\nonumber\\  
{\rm Re}M_{11} & =
& \frac{{\rm Re}(C^{\rm }g)_{11}{\rm Re}({\det A}) + {\rm
Im}(C^{\rm }g)_{11}{\rm Im}({\det A})}{\left[{\rm Re}(\det
A)\right]^{2} +
\left[{\rm Im}({\det A})\right]^{2}}f(x^{\prime})f(x)\, ,
\label{inelimmremm}
\end{eqnarray}
where
\begin{eqnarray}
{\rm Re}(C^{\rm T}g)_{11} & = & g_{11}(1 - g_{33}{\cal G}_{3}) +
g_{13}^{2}{\cal G}_{3}  - D {\rm Re}{\cal G}_{2}\nonumber \\ 
{\rm Im}(C^{\rm T}g)_{11} & = & -D{\rm Im}{\cal
G}_{2}\nonumber\\ 
{\rm Re}(\det A) & = & 1 - g_{33}{\cal G}_{3} -
g_{11}{\rm Re}{\cal G}_{1} - g_{22}{\rm Re}{\cal G}_{2} \nonumber\\
& & + (g_{11}g_{33}
- g_{13}^{2}){\cal G}_{3}{\rm Re}{\cal G}_{1} + (g_{22}g_{33} -
g_{23}^{2}){\cal G}_{3}{\rm Re}{\cal G}_{2} \nonumber \\
 & & + (g_{11}g_{22} - g_{12}^{2})({\rm Re}{\cal G}_{1} {\rm Re}{\cal
G}_{2} - {\rm Im}{\cal G}_{1}{\rm Im}{\cal G}_{2})\nonumber \\
 & & - \det g\;{\cal G}_{3}({\rm Re}{\cal
G}_{1}{\rm Re}{\cal G}_{2} - {\rm Im}{\cal G}_{1}{\rm Im}{\cal
G}_{2})\nonumber \\ 
{\rm Im}(\det A) & = & -g_{11}{\rm Im}{\cal G}_{1}
- g_{22}{\rm Im}{\cal G}_{2} + (g_{11}g_{33} - g_{13}^{2}){\cal
G}_{3}{\rm Im}{\cal G}_{1} + (g_{22}g_{33} - g_{23}^{2}){\cal
G}_{3}{\rm Im}{\cal G}_{2}\nonumber \\
 & & + (g_{11}g_{22} - g_{12}^{2})({\rm Re}{\cal G}_{1} {\rm Im}{\cal
G}_{2} + {\rm Im}{\cal G}_{1}{\rm Re}{\cal G}_{2})\nonumber \\
 & & - \det g\;{\cal G}_{3}({\rm Re}{\cal
G}_{1}{\rm Im}{\cal G}_{2} + {\rm Im}{\cal G}_{1}{\rm Re}{\cal
G}_{2})\, , \label{imrenumden}
\end{eqnarray}
and 
$$D=\left[ g_{11}g_{22}(1 - g_{33}{\cal G}_{3}) + g_{11}g_{23}^{2}
{\cal G}_{3} - g_{12}(1 - g_{33}{\cal G}_{3}) -
2g_{12}g_{13}g_{23}{\cal G}_{3} + g_{13}^{2}g_{22}{\cal G}_{3}\right]
\, , $$
and
$$ \det g=g_{11}g_{22}g_{33} - g_{11}g_{23}^{2} - g_{12}^{2}g_{33} + 
2g_{12}g_{13}g_{23} - g_{13}^{2}g_{22}\, .$$

%
%
\subsection{The interaction current contribution to the deuteron 
form  factor}
\label{examination}

We have already derived the specific formula for the three-member 
tower deuteron form factor in Section 
\ref{specific}.  The analytic form of the interaction current 
contribution, specifically in the limit as $Q^2$ approaches infinity,
merits closer examination.  It shall be shown  that this limit 
reveals a rather unusual property of the interaction current
contribution: it is constant as $Q^2$ becomes very large.

\begin{figure}[t]
\begin{minipage}{4.0in}
\begin{center}
\mbox{
   \epsfxsize=3.5in
\epsfbox{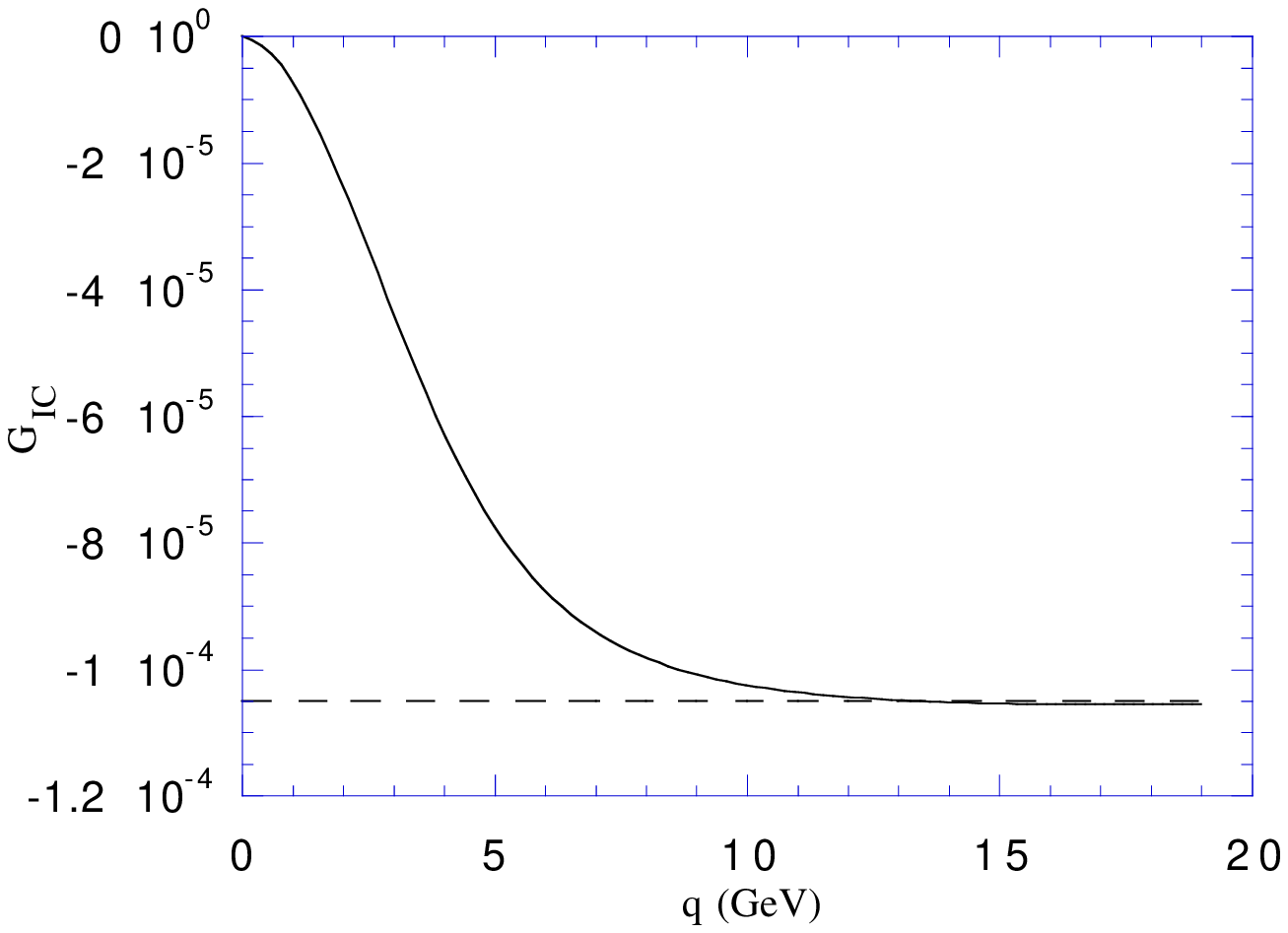}
}
\end{center}
\caption{The interaction current contribution to the deuteron form 
factor in the case of Model 1, plotted out to very large values of 
$Q$.  The dashed line shows the calculated value of the constant
which the interaction current should approach as $Q$ becomes large.}
\label{lqicc}
\end{minipage}
\end{figure}

Recall from Eq.~(\ref{tdff}) that the interaction current
contribution to the deuteron form factor is
\begin{eqnarray}
G_{\rm IC}(Q^2)=&& 2 g_{ij}\; C_i \int\frac{d^{3}p_{n}}{(2\pi)^{3}2
E} 
\phi_{j}^{B}(p_n,D)
\overline{{\cal Z}}\left(\frac{f\left[x(p_{n},
D^{\prime})\right] - f\left[x(p_{n}, D)\right]}{x(p_{n},
D^{\prime}) - x(p_{n}, D)}\right) \nonumber\\
=&& 2 g_{ij}\; C_i \;{\cal I}_j(Q^2)\, , 
\label{fint}
\end{eqnarray}
where $\overline {\cal Z}$ and $C_i$ are as defined 
in Eqs.~(\ref{zbar}) and (\ref{ci}).  Using the fact that
$\phi^B_j(p_n,D) = {\cal N}G_j(p)f\left[x(p_n,D)\right]$ [recall
Eq.~(\ref{wfeq})], Eq.~(\ref{fint}) gives the following expression
for ${\cal I}_j(Q^2)$ 
\begin{equation}
{\cal I}_j(Q^2) = {\cal N}\int\frac{d^{3}p_{n}}{(2\pi)^{3}2 E}\; 
G_j(p)f\left[x(p_n,D)\right] \overline{{\cal
Z}}\left(\frac{f\left[x(p_{n}, D^{\prime})\right] - f\left[x(p_{n},
D)\right]}{x(p_{n}, D^{\prime}) - x(p_{n}, D)}\right)\, ,
\label{redfint}
\end{equation}
This integral can be evaluated in any frame (it is covariant) and
it is convenient to evaluate it in the {\em incoming} deuteron
rest frame. In order to express the functions in
${\cal I}_j$ in the incoming deuteron rest frame, all of the 
momenta must be boosted from the
Breit frame to this new frame.  This boosting can be
accomplished by the use of the matrix $\Lambda$, which carries $D$
to rest:
\begin{equation} \Lambda \left(\begin{array}{r} D_{0} \\  \\
-{\textstyle\frac{1}{2}}{ q}
\end{array}\right) =\frac{1}{M_D} \left(\begin{array}{rrr}
D_{0}  & & {\textstyle\frac{1}{2}}{q} \\ & & \\
{\textstyle\frac{1}{2}}{q} & & D_{0}
\end{array}\right) \left(\begin{array}{r}
D_{0} \\  \\
-{\textstyle\frac{1}{2}}{ q}
\end{array}\right) = \left(\begin{array}{c}
M_{D} \\  \\   0
\end{array}\right)\,  ,
\label{boost}
\end{equation}
where $q$ is the magnitude of ${\bf q}$ (and will be taken to be in
the $\hat z$ direction).  Using this boost $\Lambda$, the outgoing
deuteron momentum becomes
\begin{equation}
\frac{1}{M_D} \left(\begin{array}{rrr}
D_{0}  & & {\textstyle\frac{1}{2}}{ q} \\ & & \\
{\textstyle\frac{1}{2}}{ q} & & D_{0}
\end{array}\right)\left(\begin{array}{r}
D_{0} \\  \\
{\textstyle\frac{1}{2}}{ q}
\end{array}\right) = \frac{1}{M_D}\left(\begin{array}{c}
M_{D}^2 + {\textstyle\frac{1}{2}}{ q}^2 \\  \\  { q} D_0  
\end{array}\right) \, ,
\label{bidm}
\end{equation}
and $x(p_{n},D) = -{p}_{n}^{{2}}$ which is clearly independent of
$q$.  Therefore the wave function $\phi^B_j(p_n,D)$ is also
independent of $q$.  However, 
\begin{eqnarray}
x(p_{n},D') =&& m_{o}^{2} -
\frac{1}{M_{D}^{2}}\left[E\left(M_{D}+\frac{{q}^{2}}{2M_{D}}\right)  
- \frac{D_0\,p_{n}qz}{M_{D}}\right]^2\nonumber\\
{\longrightarrow}&&\;  
-\frac{{q}^{4}}{4M_{D}^{4}}(E - {p}_{n}z)^{2}\quad {\rm as}\; q
\to\infty\, ,
\label{xpnpd}
\end{eqnarray}
where $z = \cos\theta$, with $\theta$ the angle between ${\bf
p}_n$ and $\hat z$.  Hence   
\begin{equation}
\frac{f\left[x(p_{n},D^{\prime})\right] -
f\left[x(p_{n},D)\right]}{x(p_{n},D^{\prime}) -
x(p_{n},D)} \longrightarrow
\frac{4M_D^4f\left[x(p_{n},D)\right]}{q^{4}
(E - {p}_{n}z)^{2}}\, ,
\label{calfqinf}
\end{equation}
because $f(x)\to0$ as $x\to\infty$. Similarily, $\overline{\cal
Z}$ becomes
\begin{eqnarray}
\overline{\cal Z} =&& \frac{E^2}{2M_D^2} + \frac{1}{2M_D^2}\left[
E\left(1 + {q^2\over 2M_D^2}\right) - {D_0\over M_D^2}\, p_n q z
\right]^2 -\frac{\left[{2D_0^2 E} -
{D_0{p}_{n}q z}\right]^2}{4M_{D}^{4}D_0^2}\nonumber\\
 {\longrightarrow}  &&
\frac{q^{4}}{8M_{D}^{6}}(E - { p}_{n}z)^{2} \quad {\rm as}\; q
\to\infty \,  .\label{calzqinf}
\end{eqnarray}
Thus the large {\it q}-dependencies of Eqs.~(\ref{calfqinf}) and
(\ref{calzqinf}) cancel, and the limit of ${\cal I}$ as
$q\to\infty$ approaches a constant
\begin{eqnarray}
{\cal I}_j(Q^2)\longrightarrow&& {\cal
N}\int\frac{d^{3}p_{n}}{(2\pi)^{3}2 E}\; 
G_j(p)f\left[x(p_n,D)\right]
\frac{q^{4}}{8M_{D}^{6}}(E - { p}_{n}z)^{2}
\frac{4M_D^4f\left[x(p_{n},D)\right]}{q^{4}
(E - {p}_{n}z)^{2}}\nonumber\\
=&& {{\cal N}\over2M_D^2}\int\frac{d^{3}p_{n}}{(2\pi)^{3}2 E}\;
G_j(p)f^2\left[x(p_n,D)\right] \nonumber\\
=&&-{{\cal N}\over2M_D^2}\,{\cal G}_j
\, , \label{constI}
\end{eqnarray}
\narrowtext
where ${\cal G}_j$ is evaluated at $D^2=M_D^2$. This means that the
interaction current part of the form factor also goes to a constant.
Calculating this constant from Eqs~(\ref{ci}) and (\ref{constI})
gives a value of $-1.12 \times 10^{-4}$, which agrees very well with
the numerical results obtained for the interaction current
contribution at very large $Q$, as shown in
Fig.~\ref{lqicc}.  We believe that this rather unusual feature of
the interacton current comes from the fact that we are using a
point-like  four-point interaction as the kernel.  At large $Q$ the
separable form factor does not prevent the basic point-like
structure we have chosen from being seen.  

%

This concludes the discussion of the specifics of the tower models.

%
%
%
%

\acknowledgments

We greatfully acknowledge the support of the Department of Energy
through grant No. DE-FG02-97ER41032.

%
%
%
\appendix
%
%
\section*{Derivation of the interaction current}
\label{derintcur}

In this appendix we derive the 
interaction current given in Eq.~(\ref{intcurr}).   While the method is
similar to that used in Ref.~\cite{ibg91}, there are certain
differences which will be detailed here.

The interaction current will be obtained by minimal substitution,
which is most easily done in coordinate space. The interaction
is chosen to depend on $D$ and $p_{n}$; the coordinates
conjugate to these momenta will be labeled $r$ and $r_d$.  From
\begin{equation}
p\cdot r + p_{n}\cdot r_{n} = D\cdot r + p_{n}\cdot r_{d}\, ,
\label{conjmom}
\end{equation}
it follows that $r_{d} = r_{n} - r$.  Using these conjugate 
coordinates, the strong interaction in  coordinate space can be
written as (factors of $g$ are neglected for the moment)
\widetext
\begin{eqnarray}
\Delta & = &
\int\frac{d^{4}p_{n}^{\prime}\,d^{4}p_{n}\,d^{4}D'\,d^{4}D
}{(2\pi)^{12}}\;e^{i(D^{\prime}\cdot r^{\prime} + p_{n}^{\prime}\cdot
r_{d}^{\prime})}e^{-i(D\cdot r + p_{n}\cdot
r_{d})}f(x^{\prime})f(x)\,\delta^{4}(D^{\prime} - D) \nonumber\\
& = & \int\frac{d^{4}p_{n}^{\prime}\,d^{4}p_{n}\,d^{4}D}
{(2\pi)^{12}}\;e^{i (D\cdot y +
p_{n}^{\prime}\cdot r'_{d}- p_{n}\cdot
r_{d})}f(x^{\prime})f(x)\nonumber\\
 & = & \Delta(y;r_{d}^{\prime},r_{d}),
\label{coordspint}
\end{eqnarray}
where $y =r^{\prime} - r$, and $x^{\prime}$ and {\it x} were defined
in Eq.~(\ref{ex}) (with $x'$ obtained from $x$ by replacing $p_n$ 
with $p'_n$).  Note that in coordinate space, the  strong interaction
is {\em not} separable, as it is in momentum  space.  

Particle 2 is neutral, so dependences of the potential
on $p_n$ and $p'_n$
\begin{equation}
p^\mu_n=-i\frac{\partial}{\partial r_{d\;\mu}}\qquad
p'^\mu_n=-i\frac{\partial}{\partial r'_{d\;\mu}}
\end{equation}
do not generate any current.  Particle 1 is charged, and
the  coordinate associated with it is $y$.  Therefore, from
minimal substitution the dependences on $D^{\mu}$  generate
\begin{equation}
D^{\mu} = -i\frac{\partial}{\partial y_{\mu}} \longrightarrow -i
\frac{\partial}{\partial y_{\mu}} + e\,\epsilon^{\mu}e^{iq\cdot y}.
\label{minsub}
\end{equation}
Applying this to Eq.~(\ref{coordspint}) generates an interaction
current proportional to $\Gamma_\mu$ 
\begin{equation}
\Delta(y;r_{d}^{\prime},r_{d}) \longrightarrow 
\Delta(y;r_{d}^{\prime},r_{d}) +
e\,\epsilon^{\mu}\,\Gamma_{\mu}(y;r_{d}^{\prime},r_{d})\, ,
\label{gammadef}
\end{equation}
where $e\,\epsilon^{\mu}\,\Gamma_\mu$ is the first order (in the
charge $e$) change to $\Delta$ which arises from the substitution
(\ref{minsub}). To find $\Gamma$, this change must be evaluated from
\begin{eqnarray}
\delta\Delta & = & e\,\epsilon^{\mu}\,
\Gamma_{\mu}(y;r_{d}^{\prime},r_{d})\nonumber\\ 
& = &
\int_{D}\Biggl(\left\{f\left[x\left(p'_n,-i\frac{\partial}{\partial
y_{\mu}} + e\,\epsilon^{\mu}\,e^{iq\cdot
y}\right)\right]f\left[x\left(p_n,-i\frac{\partial}
{\partial y_{\mu}} + e\,\epsilon^{\mu}\,e^{iq\cdot
y}\right)\right]\right\} \nonumber\\
 & & - \left\{f\left[x\left(p'_n,-i\frac{\partial} {\partial
y}\right)\right]f\left[x\left(p_n, -i\frac{\partial}
{\partial y}\right)\right]\right\}\Biggr)\;e^{i(D\cdot y +
p_{n}^{\prime}\cdot r_{d}^{\prime} - p_{n}\cdot r_{d})}\, ,
\label{deltdiff}
\end{eqnarray}
where
\begin{equation}
\int_{D} = \int\frac{d^{4}Dd^{4}p_{n}^{\prime}d^{4}p_{n}}
{(2\pi)^{12}}\, .
\label{intintintd}
\end{equation}

To evaluate this, assume that $f$ can be expanded in a Taylor
series
\begin{equation}
f(x) = \sum_{n}c_{n}\,x^{n}\, ,
\label{texpf}
\end{equation}
define ${\cal Z}_{\mu}$ by
\begin{equation}
x\left(p_n, -i\frac{\partial}{\partial y_{\mu}} + e\epsilon^{\mu}
e^{iq\cdot y}\right) = \tilde{x} =
x\left(p_n, -i\frac{\partial}{\partial y_{\mu}}\right) +
e\,\epsilon^{\mu}\,{\cal Z}_{\mu}\,e^{iq\cdot y} + {\cal O}(e^{2})
\, , \label{minsubx}
\end{equation}
and work out the difference $\tilde{x}^n-x^n$ arising from
the minimal substitution.  Since $\delta\,x=\tilde{x}-x$ does not
commute with $x$, this difference must first be written
\begin{equation}
\tilde{x}^n-x^n = \sum_{m}x^{n - 1 - m}\left(p_n,
-i\frac{\partial}{\partial y_{\mu}}\right) \,(\delta\,x)\,
x^{m}\left(p_n, -i\frac{\partial}{\partial y_{\mu}}\right)\, .
\label{xdiffaa}
\end{equation}
Now take into account the fact that each appearance of
$\partial/\partial y$ will act on {\it all\/} factors of {\it y}  to
the right of it.  This includes the factor of $\exp i(D\cdot y +
p_{n}^{\prime}\cdot r_{d} - p_{n}\cdot r_{d})$ in the definition of
$\delta\Delta$ where this expansion of the form factor will
eventually be used.   Therefore, factors of {\it x} occurring to the
right of the factor $\delta\,x$ will depend on 
{\it D} only, while factors of {\it x} occurring to the left of the
$\delta\,x$ term will  depend on $D + q = D^{\prime}$ once the
expansion is inserted into Eq.~(\ref{deltdiff}).  Hence the change in
$x^{n}$ under minimal substitution to first order in {\it e} is
\begin{eqnarray}
\tilde{x}^{n} - x^{n} =&& \sum_{m}x^{n - 1 - m}(p_{n}, D + q )
\left[e\epsilon^{\mu}{\cal Z}_{\mu}e^{iq\cdot y}\right] 
x^{m}(p_{n}, D) \nonumber\\ 
=&& e\,\epsilon^{\mu}\,{\cal Z}_{\mu}\,e^{iq\cdot y}\;
\sum_{m} x^{n - 1 - m}(p_{n}, D + q )x^{m}(p_{n}, D) \, .
\label{xdiff}
\end{eqnarray}
Using the algebraic identity
\begin{equation}
\sum_{m = 0}^{n - 1}a^{n - 1 - m}b^{m} = \frac{a^{n} - b^{n}}{a - b}
,\label{algid}
\end{equation}
Eq.~(\ref{xdiff}) becomes
\begin{equation}
\tilde{x}^{n} - x^{n} = e\,\epsilon^{\mu}\,{\cal Z}_{\mu}\,
e^{iq\cdot y}\;
\frac{x^{n}(p_{n}, D + q) - x^{n}(p_{n},D)}{x(p_{n},D + q) -
x(p_{n},D)}\, .
\label{redxdiff}
\end{equation}
Therefore, the change in {\it f} under minimal substitution is
\begin{equation}
\delta f = e\,\epsilon^{\mu}\,{\cal Z}_{\mu}\,e^{iq\cdot y}\;
\frac{f\left[x(p_{n},D+q)\right] -
f\left[x(p_{n},D)\right]}{x(p_{n},D+q) - x(p_{n},D,)}.
\label{fdiff}
\end{equation}

Note that Eq.~(\ref{deltdiff}) has the form $\tilde{f}^{\prime}
\tilde{f} - f^{\prime}f$,  where the tilde indicates the use of 
minimal substitution.  To first order in {\it e} this can be
written 
\begin{eqnarray}
\tilde{f}^{\prime}\tilde{f} - f^{\prime}f & = &
\tilde{f}^{\prime}\tilde{f} - f^{\prime}\tilde{f} + 
f^{\prime}\tilde{f} - f^{\prime}f \nonumber\\
 & = & \delta f^{\prime}\tilde{f} + f^{\prime}\delta f \nonumber\\
 & = & \delta f^{\prime}f + f^{\prime}\delta f\, .\label{fstuff}
\end{eqnarray}
Thus, Eq.~(\ref{deltdiff}) becomes
\begin{eqnarray}
\delta\Delta & = & \int_{D}e^{i\left[D'\cdot y + p_{n}^{\prime}
\cdot r_{d}^{\prime} - p_{n}\cdot
r_{d}\right]}\Biggl\{e\,\epsilon^{\mu}\,{\cal
Z}^{\prime}_{\mu}\;\frac{f\left[x(p_{n}^{\prime},D')\right] -
f\left[x(p'_{n}, D)\right]}{x(p_{n}^{\prime},D')
- x(p'_{n},D)}f\left[x(p_{n},D)\right]
\nonumber \\
 & & + e\,\epsilon^{\mu}\,{\cal Z}_{\mu}\;
f\left[x(p_{n}^{\prime},D')\right]\frac{f\left[x(p_{n},D')\right]
- f\left[x(p_{n},D)\right]}{x(p_{n},D') -
x(p_{n},D)}\Biggr\}\, .\label{findeltdiff}
\end{eqnarray}
Recalling that $\delta\Delta = e\,\epsilon^{\mu}\,\Gamma_{\mu}$, we 
can unfold the integral to give
\begin{eqnarray}
j_{\mu}^{INT} = e\,g\Biggl\{&&{\cal Z}^{\prime}_{\mu}\;
\frac{f\left[x(p_{n}^{\prime},D')\right] -
f\left[x(p'_{n}, D)\right]}{x(p_{n}^{\prime},D')
- x(p'_{n},D)}f\left[x(p_{n},D)\right]\nonumber\\
&& + {\cal
Z}_{\mu}\,f\left[x(p_{n}^{\prime},D')\right]
\frac{f\left[x(p_{n},D')\right] -
f\left[x(p_{n},D)\right]}{x(p_{n},D') - x(p_{n},D)}\Biggr\}\, ,
\label{intcurrdef}
\end{eqnarray}
which is the one-member tower version of Eq.~(\ref{intcurr}).

To complete the calculation we must find ${\cal Z}^\mu$.  We begin
from its  definition in Eq.~(\ref{xdiff}) and construct it
directly.  Using the definition of $x$, Eq.~(\ref{ex}), assuming
that $\tilde{x}$ may be expanded in a power series in $e$, using the
notation 
\begin{equation}
\tilde{D}^\mu= -i{\partial\over\partial y_\mu} + e\,\epsilon^{\mu}
\,e^{iq\cdot y} = -i{\partial\over\partial y_\mu} + e\,\delta^\mu\, , 
\end{equation}
and  keeping only those terms which are first order in
$e$, the change in  $x$ due to minimal substitution is
\begin{eqnarray}
\frac{\tilde{x} - x}{e} &=& -{(p_n\cdot \tilde{D})^2\over
e\,\tilde{D}^2} + {(p_n\cdot {D})^2\over e\, {D}^2}\nonumber\\
& = & -\left\{
{1\over 2 D^2}\;,\;\left[(p_n\cdot {D})\;(p_n\cdot \delta) +
(p_n\cdot \delta)\; (p_n\cdot {D})\right] \right\}
\nonumber\\
&  & + \left\{\lambda\;(p_{n}\cdot D)^{2}\; ,\;\left[\frac{1}{4D^{4}}
\left\{\delta, D\right\} + \left\{\delta,
D\right\} \frac{1}{4D^{4}}\right] \right\} \nonumber\\ 
& & + \left\{(1 -\lambda)(p_{n}\cdot D)^{2}\; ,\;\frac{1}{2D^{2}}
\left\{\delta, D\right\} \frac{1}{D^{2}}\right\}\, ,\label{symmxdiff}
\end{eqnarray}
where $\{A,B\}=AB+BA$, and 
\begin{equation}
\left\{\delta, D\right\}= \left\{\epsilon\cdot
D,e^{iq\cdot y}\right\}= \epsilon\cdot D\,e^{iq\cdot y} + 
e^{iq\cdot y}\,\epsilon\cdot D
\end{equation}
is the summetrized expansion of $\tilde{D}^2-D^2$.
We have expanded these functions symmetrically, using anticommutators
of non-commuting operators, since every factor of $D$ contains a
derivative which may not necessarily commute with all of its
surrounding terms.  The parameter $\lambda$ is introduced to provides a
continuous choice between two different ways of symmetrizing the
expansion of the $D^{-2}$ term; it is not clear
{\it a priori\/} which of this continuous range of choices is correct. 
(If the operators all commuted, the result would be independent of
the choice of $\lambda$.)  Extracting the common factor of
$\epsilon^{\mu}\,e^{iq\cdot y}$ [remembering that factors of $D$
contain derivatives which act all the way to the right when they are
substituted into Eq.~(\ref{deltdiff})], gives 
\begin{eqnarray}
\frac{\tilde{x} - x}{e} & = & -\epsilon^{\mu}e^{iq\cdot y}
\left\{p_{n_{\mu}}\;\;\left[p_{n}\cdot
(D+D^{\prime})\right] \left[\frac{1}{2D^{\prime^{2}}} +
\frac{1}{2D^{2}}\right]\right\}\nonumber\\
 & & + \epsilon^{\mu}e^{iq\cdot y}\Biggl\{(D+D^{\prime})_{\mu}
\left[(p_{n}\cdot D)^{2} + (p_{n}\cdot
D^{\prime})^{2}\right] \nonumber\\
& &
\qquad\qquad \times\left[\frac{\lambda}{4}\left(\frac{1}{D^{4}}  
+ \frac{1}{D^{\prime^{4}}}\right) + \frac{1 -\lambda}{2
D^{2}D^{\prime^{2}}}\right]\Biggr\}\, .\label{finxsymm}
\end{eqnarray}
Therefore, 
\begin{eqnarray}
{\cal Z}^{\mu} & = & -p_n^\mu\,\,\left[p_{n}\cdot 
(D+D^{\prime})\right]\left(\frac{1}{2 D^{\prime^{2}}} +
\frac{1}{2 D^{2}}\right) \nonumber \\
 & & + (D+D^{\prime})^{\mu}\left[(p_{n}\cdot D)^{2}  +
(p_{n}\cdot D^{\prime})^{2}\right]\left[\frac{\lambda}{4}\left(\frac{1}{D^{4}} 
+ \frac{1}{D^{\prime^{4}}}\right) + \frac{1
-\lambda}{2 D^{2}D^{\prime^{2}}}\right]\, .\label{z}
\end{eqnarray}

All that remains is to determine $\lambda$.  To do this, look at 
Eq.~(\ref{z}) in the case of elastic scattering ($D^{2} =
D^{\prime^{2}} = M_{D}^{2}$).  Contracting with $q$ gives
\begin{equation}
q_{\mu}{\cal Z}^{\mu} =
 -\left[(p_{n}\cdot D^{\prime})^{2} - 
(p_{n}\cdot D)^{2}\right]\frac{1}{M_{D}^{2}} = x(p_{n},D^{\prime}) 
- x(p_{n},D)\, .
\label{elscattqz}
\end{equation}
This is the result needed for the total current to be conserved [this
result leads directly to Eq.~(\ref{qincurr})].  Hence, current
conservation requires that this should be the result even when  the
scattering is not elastic.  In that case,
\begin{eqnarray}
 q_{\mu}{\cal Z}^{\mu} = && (p_{n}\cdot D^{\prime})^{2}
\left[-\frac{1}{2D^{\prime^{2}}}\left(2 - {3\over2}\lambda \right) -
\frac{1}{2D^{2}}\left( {3\over2} \lambda \right) + \frac{\lambda
D^{\prime^{2}}}{4D^{4}} -
\frac{\lambda D^{2}}{4D^{\prime^{4}}}\right]\nonumber\\
&& + (p_{n}\cdot D)^{2} \left[\frac{1}{2D^{2}}\left(2 - {3\over2} 
\lambda\right) + \frac{1}{2D^{\prime^{2}}}\left( {3\over2}
\lambda \right) + \frac{\lambda
D^{\prime^{2}}}{4D^{4}} -
\frac{\lambda D^{2}}{4D^{\prime^{4}}}\right] \label{inelscattqz}
\end{eqnarray}
\narrowtext
Clearly the choice of $\lambda = 0$ will reduce the expression to 
the desired form, ${\cal Z}_{\mu}$ reduces to Eq.~(\ref{zee}), and
the  interaction current (\ref{intcurrdef}) is completely specified.
The generalization of the interaction current to the case with more
than one tower member is straightforward; the result is given in
Eq.~(\ref{intcurr}).

%
%
%
%


\begin{references}
%
%
%

\vspace*{-0.4in}

\bibitem[*]{present} Based on work completed by K.~Herbst in partial
fulfilment of requirements for the Ph.D. degree at the College of
William and Mary.
%
\bibitem{r68} R. Reid, Ann. Phys. (N.Y.) {\bf 50}, 411 (1968).
%
\bibitem{mt69}R. A. Malfliet and J. A. Tjon, Nucl. Phys.  {\bf
A127}, 161 (1969).
%
\bibitem{clopz75}L. \v{C}repin\v{s}ek, C. B. Lang, H. Oberhummer, 
W. Plessas, and H. F. K. Zingl, Acta Phys. Austriaca {\bf 42}, 139
(1975).
%
\bibitem{sk80}R. R. Silbar and W. M. Kloet, Nucl. Phys. {\bf A338}, 
317 (1980).
%
\bibitem{mps82} L. Mathelitsch, W. Plessas, and W. Schweiger, Phys. 
Rev. C {\bf 26}, 65 (1982).
%
\bibitem{shf86} K. Schwarz, J. Haidenbauer, and J. Fr\"{o}hlich, 
Phys. Rev. C {\bf 33}, 456 (1986).
%
\bibitem{sp90}W. A. Schnizer and W. Plessas, Phys. Rev. C {\bf 41}, 
1095 (1990).
%
\bibitem{rt92}G. Rupp and J. A. Tjon, Phys. Rev. C {\bf 45}, 2133 
(1992).
%
\bibitem{gvoh92} F. Gross, J. W. VanOrden, and K. Holinde, Phys. 
Rev. C {\bf 45}, 2094 (1992).
%
%
\bibitem{ft75nc}J. Fleischer and J. A. Tjon, Nucl. Phys {\bf B84}, 
375 (1975); Phys. Rev. D {\bf 15}, 2537 (1977); {\bf 21}, 87 (1980).
%
\bibitem{zt80nc}M. J. Zuilhof and J. A. Tjon, Phys. Rev. C {\bf 22},
2369 (1980); {\bf 24}, 736 (1981).
%
\bibitem{ft83nc}E. vanFaassen and J. A. Tjon, Phys. Rev. C {\bf 28},
2354 (1983); {\bf 30}, 285 (1984).
%
%
\bibitem{ft86}E. vanFaassen and J. A. Tjon, Phys. Rev. C {\bf 33}, 
2105 (1986).
%
%
\bibitem{rt88}G. Rupp and J. A. Tjon, Phys. Rev. C {\bf 37}, 1729 
(1988).
%
%
\bibitem{sb51}E. E. Salpeter and H. A. Bethe, Phys. Rev. {\bf 84}, 
1232 (1951).
%
%
\bibitem{bs66}R. Blankenbecler and R. Sugar, Phys. Rev. {\bf 142},
1051 (1966).
%
%
\bibitem{g69}F. Gross, Phys. Rev. {\bf 186}, 1448 (1969).
\bibitem{g74}F. Gross, Phys. Rev. D {\bf 10}, 223 (1974).
\bibitem{g89}F. Gross, Czech. J. Phys. {\bf B 39}, 871 (1989).
\bibitem{g93book}F. Gross, {\it Relativistic Quantum Mechanics and 
Field Theory} (John Wiley and Sons, New York, 1993), Chap. 12. 
%
%
\bibitem{n93pc}V. G. J. Stoks, R. A. M. Klomp, M. C. M. Rentmeester, 
and J. J. de Swart, Phys. Rev. C {\bf 48}, 792 (1993).
%
%
\bibitem{gr87}F. Gross and D. O. Riska, Phys. Rev. C {\bf 36}, 1928 
(1987).
%
%
\bibitem{ibg91}H. Ito, W. W. Buck, and F. Gross, Phys. Rev C {\bf 43}, 
2483 (1991).
%
%
\bibitem{arbcvws83}R. A. Arndt, L. D. Roper, R. A. Bryan, R. B. Clark, 
B. J. VerWest, and P. Signell, Phys. Rev. D {\bf 28}, 97 (1983).
%
%
\bibitem{vopc}M.~Gar\c{c}on, et.al., Phys.~Rev.~C {\bf 49}, 2516
(1994).
%
%
\bibitem{gs93}F. Gross and Y. Surya, Phys. Rev. C {\bf 47}, 703 
(1993).  The formulation of the inelastic scattering  matrix and
subsequent calculations closely resemble their Appendix C.
%
%
\bibitem{t57}Y. Takahashi, Nuovo Cimento {\bf 6}, 2231 (1957).
\bibitem{w50nc}J. C. Ward, Phys. Rev. {\bf 77}, 293 (1950); {\bf 78}, 
182 (1950); Proc. Phys. Soc. {\bf 64}, 54 (1951).
%
\end{references}
\end{document}